\documentclass{llncs}
\usepackage[a4paper, margin=1.2in]{geometry}
\usepackage{graphicx}
\usepackage{bussproofs}
\usepackage{amsmath}
\usepackage{amssymb}
\usepackage{mathrsfs}
\usepackage{multicol}
\usepackage{fancyvrb}
\usepackage[utf8]{inputenc}
\usepackage[T1]{fontenc}

\usepackage{bookmark}
\usepackage{listings}
\usepackage{holindex}
\usepackage{alltt}

\newcommand{\HOLTokenStrongEquiv}{$\sim$}
\newcommand{\HOLTokenWeakEquiv}{$\approx$}
\newcommand{\HOLTokenRootedWeakEquiv}{$\approx ^c$}
\newcommand{\HOLepsilon}{$\epsilon$}

\begin{document}

\title{A Formalization of the Process Algebra CCS in HOL4}
\author{Chun Tian}
\institute{Scuola di Scienze, Universit\`{a} di Bologna\\
\email{chun.tian@studio.unibo.it}\\
Numero di matricola: 0000735539}
\maketitle

\begin{abstract}
An old formalization of the Process Algebra CCS (no value passing,
with explicit relabeling operator) has been
ported from HOL88 theorem prover to HOL4 (Kananaskis-11 and later). Transitions between CCS
processes are defined by SOS (Structural Operational Semantics)
rules, then algebaric laws for strong equivalence (including the expansion
law) were proved upon SOS transition rules.

We used HOL4's new co-inductive relation support
to re-define strong and weak bisimulation equivalances, and showed that
these new definitions are equivalent with the old ones. There's also a decision procedure for automatic
detection of CCS transitions. The aim of this project is to provide an up-to-date sound and effective
tool to support verifications and reasoning in CCS, and
to provide a formal logic basis for further theoretical developments in CCS.
\end{abstract}

\section{Introduction}

 Concurrency Theory \cite{Gorrieri:2015jt} has been successfully applied for
 explaining many concurrency
 phenomenons in Computer Science. In this theory, reactive systems can be modelled as
 (possibly infinite) directed graphs of some (atomic) states with labeled edges as
 transitions between these states. Reactive systems defined in this way
 are called \emph{Labeled Transition Systems} (LTSs) \cite{Keller:1976kl}.

 But it's usually inconvenient to describe and use
 (rooted) LTSs when studying the behaviors of reactive systems,
 especially when the system has infinite number of states.
 To overcome such difficulities, different compact
 representations were invented as languages for describing reactive
 systems. One notable is Milner's \emph{Calculus of Communicating Systems}
 (CCS) \cite{Milner:2012uz}. Such a language can be considered as
 process algebras (or syntactic calculus), with LTSs as its underlying
 semantic models.
The possible transitions of
 processes can be defined by a group of inference rules called
 Structural Operational Semantics (SOS), then these rules can be used
 as a set of axioms for proving all of algebraic laws about the
 equivalence of CCS processes.

There're usually two methods to study the behaviors of reactive systems. One is
through the behavior equivalence checking between the \emph{specification} and
\emph{implementation} of the same reactive system. Notable behavior
equivalences include the strong and (rooted) weak bisimularities.
The other method is to do model
checking directly on the implementation, in which the target
properties were usually expressed in $\mu$-Calculus or Henessy-Milner
Logics (HML)\footnote{A formulization of HML was part of the original
  work by Prof. Monica Nesi, but it's not included in this project,
  because the related HOL88 scripts were not sent to the author during
  the porting work. The porting of HML formulization is planed in futher projects.}.

Concurrency Theory is a proof-intense area: the
contents of related course and textbooks are full of theorems, definitions and
proofs, just like mathematics. On the other side, non-trivial algorithms and
their optimizations used in model checking software were usually not
considered as necessary part of textbooks. However, most
of these theorems were only done in pencil-and-paper, i.e. they have no
formal proofs, although their correctness rarely raises doubts.
When people was publishing new process algebras
like new CCS variants, usually there's no corresponding software
environments to actually play with the new theorey, nor the theorems
presented in the paper were formally verified. We think the main
reason is the lack of the formal basis in which fundamental datatypes
and relations and theorems were supplied for further extension.

The CCS formalization done by Monica Nesi in 1992-1995 was a success.
The work was ever considered as a major success story
\cite{Gordon:2000vt} of HOL theorem
prover, not to further mention that both CCS
and HOL (derived from Edinburgh LCF) were derived from the initial pioneering work by the
same scholar, Robin Milner. After more than 20 years, HOL88 and HOL90
were replaced by HOL4 (latest release is called Kananaskis-11); even
the
underlying programming language for writing HOL has changed from
Classic ML (defined on top of Common Lisp) to Standard ML. But the CCS
formalization seems being forgotten and to the best of our knowledge
there's no other CCS formalizations done in HOL\footnote{Except that Prof. Monica
  Nesi further published her formalizations of value-passing CCS in
  1994 \cite{nesi1994value} and 1997 \cite{Nesi:1997di}.} and other theorem
provers (e.g. Coq). New variants of concurrency theory and new
theorems kept being
introduced, but rarely they were formally verified. This is a rather
unacceptable situation from the view of the author. As a little
contributions to adademic, the author has spent several months porting
all the old code (provided by Prof. Nesi) to latest HOL4 with some
improvements, new theorems, also with new features (e.g. co-inductive
relation) used as alternative ways to define certain important
concepts (e.g. bisimulation equivalence) in concurrency theory.

Currently this work contains about 6000 lines of proof scripts in Standard ML\footnote{Currently it's stored in GitHub:
  \url{https://github.com/binghe/informatica-public/tree/master/CCS},
  and recently these code has entered into HOL official code base in
  its "examples/CCS" folder.}, with
examples. About 170 theorems were proved (most of them were part of the old project), including the Expansion
Law for strong equivalance. There's also a ML function for computing CCS transitions from
any given CCS process, and the output is a theorem which completely
characterizes its transitions. Programs written in this way
can be seen as kind of trusted computing, in the sense that, whenever
the computation is terminated, the result must be correct, since it's
a proven theorem. This prevented any doubts from the possible bugs
existed in the software. For instance, if there's a hidding bug in
Concurrency Workbench and have caused two complicated CCS processes
being considered as equivalent but actually they're not, there's
almost no way to know, except for comparing the results from another
different software. Of course, any program including our ML functions
could have bugs, thus we're not sure if the program can always give
the result for any valid input, but as long as it does have a result,
the result MUST be true (since it's a theorem) thus can be fully
trusted. We think this is a major advantage to use theorem provers in
place of usual programming languages (e.g. C, Java or OCaml) for implementing
software verification tools, as reliabilty is more important than other factors.

\section{Background}

Back to one year ago (May 2016), soon after the author has just attended his 2nd-year course
\emph{MODELLI E SISTEMI CONCORRENTI (Concurrent Models and Systems)}
of Computer Science (Informatica) at University of Bologna, he was
looking on Internet for connections
between CCS and the HOL theorem prover (HOL4) that he just began to
learn. At that time, the author didn't know how to use HOL4 yet, but he
liked it and the whole \emph{formal methods} area very much.

Fortunately, the author found a paper \cite{Gordon:2000vt} about the history
of HOL theorem prover, in which Mike Gordon wrote:
\begin{quote}
I had been impressed by how the Expansion Theorem of Milner's Calculus
of Communicating Systems (CCS)
 \cite{Milner:2017tw}
enabled a direct description of the behaviour of a composite agent to
be calculated
from the parallel composition of its individual components.

... Incidently, not only was CCS's Expansion Theorem an
inspirational stepping stone from LCF via LSM to HOL, but recently things
have come `full circle' and Monica Nesi has used HOL to provide proof support
for CCS, including the mechanisation of the Expansion Theorem \cite{Nesi:1992ve}.
\end{quote}

This is how the author found the paper of Prof. Monica Nesi and some deep
connections between LCF (Robin Milner), HOL (Mike Gordon) and CCS
(again Robin Milner!).  Then on May 24, 2016, the author posted on
HOL's mailing list asking for the proof scripts code mentioned in the
paper. Surprisingly Mike Gordon replied that mail with the following
contents:
\begin{quote}
``Hi,

I don't have Monica's email address, but I do have her husband's, so
I've forwarded your email to him.

Cheers,

Mike''
\end{quote}
In later mails in the same day, Mike Gordon also told the author that,
``Monica Nesi is also in Italy: she works at the University of
L'Aquila (unless she has moved recently)''.\footnote{I should have really
  searched on Internet first...} Just one day later, on May 25, 2016, Professor Nesi sent an email to the author:
\begin{quote}
``Dear student, 

I am Monica Nesi (a "she" :-)) from University of L'Aquila. 
My HOL scripts on CCS formalization are not available on Internet, but just give me 
time to find back my files and I will send them to you. I haven't been working on that 
after my PhD, already more than 20 years ago, and I am pleased that someone might be
interested in having a look and maybe do something similar in HOL4. 

I have some deadlines to meet by Friday. I will come back to you asap. 

Best regards,
Monica Nesi''
\end{quote}
And finally on June 7, the author received about 4500 lines of HOL88
proof scripts in 21 disk files. Here is a list of
these files:
\begin{verbatim}
syntax.ml
syntax_aux.ml
aux_fun.ml
basic_rule_tac.ml
opsem.ml
runM.ml
StrongEQ/basic_conv_tac.ml
StrongEQ/basic_fun.ml
StrongEQ/par_strong_laws.ml
StrongEQ/parallel_new.ml
StrongEQ/rec_strong_laws.ml
StrongEQ/relab_strong_laws.ml
StrongEQ/restr_strong_laws.ml
StrongEQ/strong_par_conv_new.ml
StrongEQ/strong_rec_conv.ml
StrongEQ/strong_relab_conv.ml
StrongEQ/strong_restr_conv.ml
StrongEQ/strong_sem.ml
StrongEQ/strong_sum_conv.ml
StrongEQ/strong_tac.ml
StrongEQ/sum_strong_laws.ml
\end{verbatim}
These include basic CCS definitions, proofs for all algebraic laws (including the
expansion law) for strong equivalence, and a complicated ML function
(in \texttt{runM.ml}) for automatically computing the transitions from
a given CCS process.

The author saved these code and continued learning HOL4. Finally,
starting from Jan 2017, the author was able to read proof scripts
written in HOL4 and prove some new theorems in it. With several
theorem proving projects doing in parallel, and kindly help from
people of HOL
community (Thomas Tuerk, Michael Norrish, Ramana Kumar, etc.), the author quickly
improved his proof skills (in HOL4) and programing kills (in Standard
ML). With a Classic ML document \cite{Kreitz:2011tw} found on
Internet, porting Classic ML code into Standard ML seems quite
straightforward.

The porting process of old HOL88 proof scripts is not very difficult: the
underlying HOL logic didn't change at all, so is the name of almost all
tacticals and other ML functions. So basically what the author did is
the following:
\begin{enumerate}
\item Copy a piece of ML code from old files into current HOL proof
  script;
\item Change the grammar from Classic ML to Standard ML;
\item Replay the proof in HOL4's interactive proof manager;
\item Make necessary changes for those tacticals which has slightly
  changed their semantics in HOL4;
\item Save the working proof in forms of HOL \texttt{store_thm}
  function calls and go to next theorem.
\end{enumerate}

Beside those small new inventions and some new ways to define old concepts
(e.g. strong equivalance), the major efforts in this project were lots
of time spent on carefully
replaying each of the proofs: most proofs on strong laws were quite
long (usually made of hundreds of tacticals, in four or five
levels). In the time of HOL88, there's NO automatic first-order
proof searching tools like Mason and Metis, nor the Q method, thus all
proofs were done manually 
including each small steps involving only bool theory and higher order
logic (beta-conversion, etc.). And whatever any literal terms are mentioned
in the proof, full type info must be given manually, this makes the
proof longer, but each step is very clear.

On the other side, Prof. Roberto Gorrieri was informed since
the very beginning. He agreed that this
project could be used as part of the exam. Thus the current
paper (as a project report) is actually the the exam paper for the author's 
\emph{MODELLI E SISTEMI CONCORRENTI (Concurrent Models and Systems)} course.

\section{A Formalization of CCS in HOL4}

The precise class of CCS we have formalized here, is
$\mathrm{CCS}^{\mathrm{rel}}$, the Fininary CCS with explicit relabeling
operator.\footnote{The relabeling operator in core CCS syntax is
  not necessary, because it's possible to define the \emph{Syntactic
    relabeling} as a recursive function on top of other CCS operators.
  But we think having native relabeling facility makes many things
  easier}

\subsection{Labels and Actions}

In most literature, the concept \emph{Action} was usually defined as the
union of a countable set of input actions $\mathscr{L}$ and output
actions (co-actions) $\overline{\mathscr{L}}$ and a special invisible
action $\tau \notin \mathscr{L} \cup \overline{\mathscr{L}}$. Actions
are also called labels when LTS is considered.

In the formalization of CCS, however, it's better to have two distinct
types: the  type \HOLinline{\HOLTyOp{Label}} is the union of input and
output (visible) actions, and the data type \HOLinline{\HOLTyOp{Action}} is the
union of all visible and invisible actions.  This is better because certain
formulae and constructors in CCS doesn't accept $\tau$ as valid
actions, e.g. the restriction operator of CCS. Thus, having two
distinct types it's possible to make sure all CCS terms constructed
from all possible values of their parameters are valid. Thus we have defined the
following two data types in HOL:
\begin{lstlisting}
Datatype `Label = name string | coname string`;
Datatype `Action = tau | label Label`;
\end{lstlisting}
Noticed that, the type \HOLinline{\HOLTyOp{Action}} contains the type Label, but they're not
in sub-type relation (there's no such support in HOL): for operators
accepting \HOLinline{\HOLTyOp{Action}}, if a term of Label were used, will result into type mismatch.

Here we have used the string type provided in HOL's \texttt{stringTheory}, thus
the distinction and injectivity of the type \HOLinline{\HOLTyOp{Label}}  also depends on the
distinction and injectivity of type \HOLinline{\HOLTyOp{string}} proved in HOL's \texttt{stringTheory}.
In theory, it's possible to use type variables instead of
the string type and allow arbitrary types being used as labels,
however we found no such needs so far.

Thus, an input action $a$ must be represented as ``\HOLinline{\HOLConst{In} \HOLStringLit{a}}'' in HOL, while a output action (co-action) $\bar{a}$ must be
represented as ``\HOLinline{\HOLConst{Out} \HOLStringLit{a}}''. Instead, the invisible
action can be written either in ASCII form \texttt{tau} or Greek
letter \HOLinline{\HOLSymConst{\ensuremath{\tau}}} in Unicode. This makes literature actions quite long,
and we have defined the following syntactic
sugars\footnote{This new solution based overloading was suggested by Michael Norrish.} as compat
representations of actions (not part of the
original CCS work):
\begin{lstlisting}
val _ = overload_on ("In", ``\a. label (name a)``);
val _ = overload_on ("Out", ``\a. label (coname a)``);
\end{lstlisting}
As the results, whenever literal visible actions will appear in above
compact forms automatically:
\begin{lstlisting}
> ``label (coname "a")``;
val it = ``Out "a"``: term
> ``label (name "b")``;
val it = ``In "b"``: term
\end{lstlisting}

The main operation on the types \HOLinline{\HOLTyOp{Label}} and \HOLinline{\HOLTyOp{Action}}
is \texttt{COMPL} which gets their complements: (for convinence we
also define the complement of $\tau$ as itself)
\begin{alltt}
\HOLTokenTurnstile{} (\HOLSymConst{\HOLTokenForall{}}\HOLBoundVar{s}. \HOLConst{COMPL} (\HOLConst{name} \HOLBoundVar{s}) \HOLSymConst{=} \HOLConst{coname} \HOLBoundVar{s}) \HOLSymConst{\HOLTokenConj{}}
   \HOLSymConst{\HOLTokenForall{}}\HOLBoundVar{s}. \HOLConst{COMPL} (\HOLConst{coname} \HOLBoundVar{s}) \HOLSymConst{=} \HOLConst{name} \HOLBoundVar{s}
\HOLTokenTurnstile{} (\HOLSymConst{\HOLTokenForall{}}\HOLBoundVar{l}. \HOLConst{COMPL} (\HOLConst{label} \HOLBoundVar{l}) \HOLSymConst{=} \HOLConst{label} (\HOLConst{COMPL} \HOLBoundVar{l})) \HOLSymConst{\HOLTokenConj{}} (\HOLConst{COMPL} \HOLSymConst{\ensuremath{\tau}} \HOLSymConst{=} \HOLSymConst{\ensuremath{\tau}})
\end{alltt}
As we know \HOLinline{\HOLTyOp{Label}} and \HOLinline{\HOLTyOp{Action}} are different types,
the \texttt{COMPL} operator on them are actually overloaded operator
of \texttt{COMPL_LAB} and \texttt{COMPL_ACT}, the complement operator
for \HOLinline{\HOLTyOp{Label}} and \HOLinline{\HOLTyOp{Action}}.

The key theorem about \HOLinline{\HOLTyOp{Label}} says that, doing complements
twice for the same label gets the label itself:
\begin{alltt}
COMPL_COMPL_LAB:
\HOLTokenTurnstile{} \HOLConst{COMPL} (\HOLConst{COMPL} \HOLFreeVar{l}) \HOLSymConst{=} \HOLFreeVar{l}
\end{alltt}
There's also a similar theorem for the double-complements of
\HOLinline{\HOLTyOp{Action}}.

The following table listed the notation of various actions, with notations from
\emph{Currency Workbench} \cite{Moller:2017uq} compared:
\begin{quote}
\begin{tabular}{|c|c|c|c|c|}
\hline
\textbf{Action} & \textbf{notation} & \textbf{CWB}
  & \textbf{HOL (ASCII)} &   \textbf{HOL
                                    (compact form)}\\
\hline
internal action & $\tau$ & \texttt{tau} & \texttt{tau} & \HOLinline{\HOLSymConst{\ensuremath{\tau}}} \\
input action & $a$ & \texttt{a} & \texttt{label (name "a")} & \HOLinline{\HOLConst{In} \HOLStringLit{a}} \\
output action & $\bar{a}$ & \texttt{'a} & \texttt{label (coname "a")} & \HOLinline{\HOLConst{Out} \HOLStringLit{a}} \\
\hline
\end{tabular}
\end{quote}

\subsection{Relabeling}

In the literature, Relabeling is usually defined as an unary
substitution operator: $\_ [b/a]$ takes a unary substitution $b/a$
(hence, $a\neq b$), and a process $p$ to construct a new process
$p[b/a]$, whose semantics is that of $p$, where action $a (\bar{a})$
is turned into $b(\bar{b})$. And multi-label relabeling can be done by
appending more unary substitution operators to the new process. The
order of multiple relabelings is important, especially when new labels introduced
in previous relabeling operation were further relabeled.

In our formalization, instead we support multi-label relabeling in one
opearation, and instead of using a list of substituions, we have
defined a new fundamental type in called
\HOLinline{\HOLTyOp{Relabeling}}.\footnote{The idea of defining relabeling as type
  bijections belongs to Prof. Monica Nesi. The author did nothing but
  the porting work from HOL88 to HOL4. Fortunately the related API
  didn't change at all.} A \HOLinline{\HOLTyOp{Relabeling}} is a
abstract type which is bijected into a subset of function of type
\HOLinline{\HOLTyOp{Label} -> \HOLTyOp{Label}}, which is called the \emph{representation} of
the type \HOLinline{\HOLTyOp{Relabeling}}. Not all functions of type \HOLinline{\HOLTyOp{Label} -> \HOLTyOp{Label}} are valid representations of \HOLinline{\HOLTyOp{Relabeling}}, but only
functions which satisfy the following property:
\begin{alltt}
\HOLTokenTurnstile{} \HOLConst{Is_Relabeling} \HOLFreeVar{f} \HOLSymConst{\HOLTokenEquiv{}} \HOLSymConst{\HOLTokenForall{}}\HOLBoundVar{s}. \HOLFreeVar{f} (\HOLConst{coname} \HOLBoundVar{s}) \HOLSymConst{=} \HOLConst{COMPL} (\HOLFreeVar{f} (\HOLConst{name} \HOLBoundVar{s}))
\end{alltt}

Noticed that, any identify function of type \HOLinline{\HOLTyOp{Label} -> \HOLTyOp{Label}}
also satisfy above property. Thus, beside specific substitutions that
we want, all
relabeling functions must be able to handle all other labels too (just
return the same label as input). (As we'll see later, such
requirements could reduce the two rules for relabelling into just one).

But usually it's more convenient to represent relabeling functions as
a list of substitutions of type \HOLinline{(\HOLTyOp{Label} \HOLTokenProd{} \HOLTyOp{Label}) \HOLTyOp{list}}. The
operator \HOLinline{\HOLConst{RELAB}} can be used to define such a relabeling
function. For instance, the term \HOLinline{\HOLConst{RELAB} [(\HOLConst{name} \HOLStringLit{b}\HOLSymConst{,}\HOLConst{name} \HOLStringLit{a}); (\HOLConst{name} \HOLStringLit{d}\HOLSymConst{,}\HOLConst{name} \HOLStringLit{c})]} can be used in place of a relabeling operator
$[b/a, d/c]$, because its type is \HOLinline{\HOLTyOp{Relabeling}}. And it must be
understood that, all rebabeling functions are total functions: for all other labels except \texttt{a} and
\texttt{c}, the substitution will be themselves (another way to
express ``no relabeling'').

Finally, have the relabeling facility defined as a multi-label
relabeling function and part of CCS syntax, we can completely
avoid the complexity of the Syntactic Substitution (c.f. p.171 of
\cite{Gorrieri:2015jt}) which has a quite complicated recursive
definition\footnote{However, syntactic relabeling is still considered
  as an ``economic'' way of doing relabeling, because having one
  native CCS operator will also introduce the corresponding SOS inference rules and
  equivalence laws.} and heavily depends on some other recursive functions
like $fn(\cdot)$ (free names) and $bn(\cdot)$ (bound names) for CCS
processes (in our project, these functions are not included nor needed).

\subsection{CCS processes and operators}

The type \HOLinline{\HOLTyOp{CCS}} is defined as an inductive data type: (thus it
must be finitary)
\begin{lstlisting}
val _ = Datatype `CCS = nil
		      | var string
		      | prefix Action CCS
		      | sum CCS CCS
		      | par CCS CCS
		      | restr (Label set) CCS
		      | relab CCS Relabeling
		      | rec string CCS`;
\end{lstlisting}

In HOL4, we have added some minimal grammar support, to represent CCS
processes in more compact forms (not available in HOL88).
The following table listed the notation of typical CCS processes and
major operators supported by above definition, with notations from
\emph{Currency Workbench} \cite{Moller:2017uq} compared:

\begin{quote}
\begin{tabular}{|c|c|c|c|c|}
\hline
\textbf{op name} & \textbf{notation} & \textbf{CWB}
  & \textbf{HOL (ASCII)} &   \textbf{HOL (compact)}\\
\hline
Deadlock (nil) & $\textbf{0}$ & \texttt{0} & \texttt{nil} &
                                                            \HOLinline{\HOLConst{nil}} \\
Prefix & $a.0$ & \texttt{a.0} & \texttt{prefix (label (name "a")) nil} & \HOLinline{\HOLConst{In} \HOLStringLit{a}\HOLSymConst{..}\HOLConst{nil}} \\
Sum & $p + q$ & \texttt{p + q} & \texttt{sum p q} & \texttt{p + q} \\
Parallel & $p | q$ & \texttt{p | q} & \texttt{par p q} & \texttt{p || q} \\
Restriction of action & $(\nu a)p$ & \texttt{p \textbackslash{}  a} &
                                                                     \texttt{nu
                                                                      \{
                                                                      "a"
                                                                      \}  p}
                                  & \HOLinline{\HOLSymConst{\ensuremath{\nu}} \HOLStringLit{a} \HOLFreeVar{p}} \\
Restriction of actions & $(\nu L)p$ & \texttt{p \textbackslash{}  L} &
                                                               \texttt{restr  L p} &   \HOLinline{\HOLSymConst{\ensuremath{\nu}} \HOLFreeVar{L} \HOLFreeVar{p}}\\
\hline
\end{tabular}
\end{quote}

For Relabeling, as we described in the last section, to express $p[b/a]$,
it must be written as \HOLinline{\HOLConst{relab} \HOLFreeVar{p} (\HOLConst{RELAB} [(\HOLConst{name} \HOLStringLit{b}\HOLSymConst{,}\HOLConst{name} \HOLStringLit{a})])},
which is a little long.

For CCS processes defined by one or more constants, in our formalization
in HOL4, all constants must be written into single term. (This is necessary for
theorem proving, because otherwise there's no way to store all
information into single variable in CCS-related theorems)  The syntax
for defining new constants is \HOLinline{\HOLConst{rec}} and the syntax to actually
use a constant is \HOLinline{\HOLConst{var}}. To see how these operators are actually
used, consider the following CCS process (the famous coffee machine
model from \cite{Gorrieri:2015jt}):
\begin{align*}
VM & \overset{def}{=} coin.(\text{ask-esp}.VM_1 + \text{ask-am}.VM_2) \\
VM_1 & \overset{def}{=} \overline{\text{esp-coffee}}.VM \\
VM_2 & \overset{def}{=} \overline{\text{am-coffee}}.VM
\end{align*}

In our formalization in HOL4, the above CCS process can be represented as
the following single term:
\begin{lstlisting}
``rec "VM"
    (In "coin"
     ..
     (In "ask-esp" .. (rec "VM1" (Out "esp-coffee"..var "VM")) +
      In "ask-am" .. (rec "VM2" (Out "am-coffee"..var "VM"))))``
\end{lstlisting}
That is, for the first time a new constant appears, use \HOLinline{\HOLConst{rec}}
with the name of constants as string to ``declare'' it; when any
constant appears again, use \HOLinline{\HOLConst{var}} to access it.

Finally, although not part of the formal definition, the
\textbf{if-then-else} construct from value-passing CCS is automatically supported by
HOL. This is because, for any boolean value $b$ and two terms $t_1$ and $t_2$ of type
$\alpha$, the term \HOLinline{\HOLKeyword{if} \HOLFreeVar{b} \HOLKeyword{then} \HOLFreeVar{t\sb{\mathrm{1}}} \HOLKeyword{else} \HOLFreeVar{t\sb{\mathrm{2}}}} has also the type
$\alpha$. Thus the conditional term can legally appears inside other
CCS processes as a sub-process. We'll see in next section that it's
necessary for handling transitions of CCS processes containing
constants.\footnote{The other similar benefit from HOL is the \textbf{let-in} binding
construct, but so far it's not well supported.}

\subsection{CCS transitions and SOS inference rules}

The transitions of CCS processes were defined by the following
Structural Operational Semantics (SOS for short) rules:
\begin{multicols}{2}

\begin{prooftree}
\AxiomC{}
\LeftLabel{(Perf)}
\UnaryInfC{$\mu.p \overset{\mu}\longrightarrow p$}
\end{prooftree}

\begin{prooftree}
\AxiomC{$q[\text{\texttt{rec} } x. q\enspace / \enspace x] \overset{\mu}\longrightarrow r$}
\LeftLabel{(Rec)} 
\UnaryInfC{$\text{\texttt{rec} } x. q \overset{\mu}\longrightarrow r$}
\end{prooftree}

\begin{prooftree}
\AxiomC{$p \overset{\mu}\longrightarrow p'$}
\LeftLabel{($\text{Sum}_1$)}
\UnaryInfC{$p + q \overset{\mu}\longrightarrow p'$}
\end{prooftree}

\begin{prooftree}
\AxiomC{$q \overset{\mu}\longrightarrow q'$}
\LeftLabel{($\text{Sum}_2$)}
\UnaryInfC{$p + q \overset{\mu}\longrightarrow q'$}
\end{prooftree}

\begin{prooftree}
\AxiomC{$p \overset{\mu}\longrightarrow p'$}
\LeftLabel{($\text{Par}_1$)}
\UnaryInfC{$p | q \overset{\mu}\longrightarrow p' | q$}
\end{prooftree}

\begin{prooftree}
\AxiomC{$q \overset{\mu}\longrightarrow q'$}
\LeftLabel{($\text{Par}_2$)}
\UnaryInfC{$p | q \overset{\mu}\longrightarrow p | q'$}
\end{prooftree}

\begin{prooftree}
\AxiomC{$p \overset{\alpha}\longrightarrow p'$}
\AxiomC{$q \overset{\bar{\alpha}}\longrightarrow q'$}
\LeftLabel{($\text{Par}_3$)} 
\BinaryInfC{$p | q \overset{\tau}\longrightarrow p' | q'$}
\end{prooftree}

\begin{prooftree}
\AxiomC{$p \overset{\mu}\longrightarrow p'$}
\LeftLabel{(Res)}
\RightLabel{$\mu \neq a,\bar{a}$}
\UnaryInfC{$(\nu a)p \overset{\mu}\longrightarrow (\nu a) p'$}
\end{prooftree}
\end{multicols}

Besides, we have a rule for relabeling:
\begin{center}
\begin{prooftree}
\AxiomC{$p \overset{\mu}\longrightarrow p'$}
\LeftLabel{(Rel)}
\UnaryInfC{$p[f] \overset{f(\mu)}\longrightarrow (p'[f]$}
\end{prooftree}
\end{center}

In \cite{Gorrieri:2015jt}, the rule $\text{Par}_3$ is called ``Com'' (rule
of communication), and
the rule ``Rec'' (in a different form based on separated agent definitions) is also called ``Cons'' (rule of constants). (Here we
have preserved the rule names in the HOL88 work, because it's easier to
locate for their names in the proof scripts.)

From the view of theorem prover (or just first-order logic), these inference rules are nothing but an \emph{inductive
  definition} on 3-ary relation \HOLinline{\HOLConst{TRANS}} (with compact
representation \texttt{--()->})  of type \HOLinline{\HOLTyOp{CCS} -> \HOLTyOp{Action} -> \HOLTyOp{CCS} -> \HOLTyOp{bool}}, generated by HOL4's function
\texttt{Hol_reln} \cite{Anonymous:Bxz1gZYL}. Then we break them into separated theorems as
primitive inference rules\footnote{They're considered as the axioms in
  our logic system, however they're not defined directly as
  axioms. HOL makes sure in such cases the logic system is still consistent.}:
\begin{alltt}
PREFIX: \HOLTokenTurnstile{} \HOLFreeVar{u}\HOLSymConst{..}\HOLFreeVar{E} --\HOLFreeVar{u}-> \HOLFreeVar{E}
REC:    \HOLTokenTurnstile{} \HOLConst{CCS_Subst} \HOLFreeVar{E} (\HOLConst{rec} \HOLFreeVar{X} \HOLFreeVar{E}) \HOLFreeVar{X} --\HOLFreeVar{u}-> \HOLFreeVar{E\sb{\mathrm{1}}} \HOLSymConst{\HOLTokenImp{}} \HOLConst{rec} \HOLFreeVar{X} \HOLFreeVar{E} --\HOLFreeVar{u}-> \HOLFreeVar{E\sb{\mathrm{1}}}
SUM1:   \HOLTokenTurnstile{} \HOLFreeVar{E} --\HOLFreeVar{u}-> \HOLFreeVar{E\sb{\mathrm{1}}} \HOLSymConst{\HOLTokenImp{}} \HOLFreeVar{E} \HOLSymConst{+} \HOLFreeVar{E\sp{\prime}} --\HOLFreeVar{u}-> \HOLFreeVar{E\sb{\mathrm{1}}}
SUM2:   \HOLTokenTurnstile{} \HOLFreeVar{E} --\HOLFreeVar{u}-> \HOLFreeVar{E\sb{\mathrm{1}}} \HOLSymConst{\HOLTokenImp{}} \HOLFreeVar{E\sp{\prime}} \HOLSymConst{+} \HOLFreeVar{E} --\HOLFreeVar{u}-> \HOLFreeVar{E\sb{\mathrm{1}}}
PAR1:   \HOLTokenTurnstile{} \HOLFreeVar{E} --\HOLFreeVar{u}-> \HOLFreeVar{E\sb{\mathrm{1}}} \HOLSymConst{\HOLTokenImp{}} \HOLFreeVar{E} \HOLSymConst{||} \HOLFreeVar{E\sp{\prime}} --\HOLFreeVar{u}-> \HOLFreeVar{E\sb{\mathrm{1}}} \HOLSymConst{||} \HOLFreeVar{E\sp{\prime}}
PAR2:   \HOLTokenTurnstile{} \HOLFreeVar{E} --\HOLFreeVar{u}-> \HOLFreeVar{E\sb{\mathrm{1}}} \HOLSymConst{\HOLTokenImp{}} \HOLFreeVar{E\sp{\prime}} \HOLSymConst{||} \HOLFreeVar{E} --\HOLFreeVar{u}-> \HOLFreeVar{E\sp{\prime}} \HOLSymConst{||} \HOLFreeVar{E\sb{\mathrm{1}}}
PAR3:
\HOLTokenTurnstile{} \HOLFreeVar{E} --\HOLConst{label} \HOLFreeVar{l}-> \HOLFreeVar{E\sb{\mathrm{1}}} \HOLSymConst{\HOLTokenConj{}} \HOLFreeVar{E\sp{\prime}} --\HOLConst{label} (\HOLConst{COMPL} \HOLFreeVar{l})-> \HOLFreeVar{E\sb{\mathrm{2}}} \HOLSymConst{\HOLTokenImp{}}
   \HOLFreeVar{E} \HOLSymConst{||} \HOLFreeVar{E\sp{\prime}} --\HOLSymConst{\ensuremath{\tau}}-> \HOLFreeVar{E\sb{\mathrm{1}}} \HOLSymConst{||} \HOLFreeVar{E\sb{\mathrm{2}}}
RESTR:
\HOLTokenTurnstile{} \HOLFreeVar{E} --\HOLFreeVar{u}-> \HOLFreeVar{E\sp{\prime}} \HOLSymConst{\HOLTokenConj{}}
   ((\HOLFreeVar{u} \HOLSymConst{=} \HOLSymConst{\ensuremath{\tau}}) \HOLSymConst{\HOLTokenDisj{}} (\HOLFreeVar{u} \HOLSymConst{=} \HOLConst{label} \HOLFreeVar{l}) \HOLSymConst{\HOLTokenConj{}} \HOLFreeVar{l} \HOLSymConst{\HOLTokenNotIn{}} \HOLFreeVar{L} \HOLSymConst{\HOLTokenConj{}} \HOLConst{COMPL} \HOLFreeVar{l} \HOLSymConst{\HOLTokenNotIn{}} \HOLFreeVar{L}) \HOLSymConst{\HOLTokenImp{}}
   \HOLSymConst{\ensuremath{\nu}} \HOLFreeVar{L} \HOLFreeVar{E} --\HOLFreeVar{u}-> \HOLSymConst{\ensuremath{\nu}} \HOLFreeVar{L} \HOLFreeVar{E\sp{\prime}}
RELAB:
\HOLTokenTurnstile{} \HOLFreeVar{E} --\HOLFreeVar{u}-> \HOLFreeVar{E\sp{\prime}} \HOLSymConst{\HOLTokenImp{}} \HOLConst{relab} \HOLFreeVar{E} \HOLFreeVar{rf} --\HOLConst{relabel} \HOLFreeVar{rf} \HOLFreeVar{u}-> \HOLConst{relab} \HOLFreeVar{E\sp{\prime}} \HOLFreeVar{rf}
\end{alltt}

Noticed that, in the rule \texttt{REC}, a recursive function
\HOLinline{\HOLConst{CCS_Subst}} was used. It has the following definition which
depends on the conditional clause (\texttt{if .. then .. else ..}):
\begin{alltt}
\HOLTokenTurnstile{} (\HOLSymConst{\HOLTokenForall{}}\HOLBoundVar{E\sp{\prime}} \HOLBoundVar{X}. \HOLConst{CCS_Subst} \HOLConst{nil} \HOLBoundVar{E\sp{\prime}} \HOLBoundVar{X} \HOLSymConst{=} \HOLConst{nil}) \HOLSymConst{\HOLTokenConj{}}
   (\HOLSymConst{\HOLTokenForall{}}\HOLBoundVar{u} \HOLBoundVar{E} \HOLBoundVar{E\sp{\prime}} \HOLBoundVar{X}. \HOLConst{CCS_Subst} (\HOLBoundVar{u}\HOLSymConst{..}\HOLBoundVar{E}) \HOLBoundVar{E\sp{\prime}} \HOLBoundVar{X} \HOLSymConst{=} \HOLBoundVar{u}\HOLSymConst{..}\HOLConst{CCS_Subst} \HOLBoundVar{E} \HOLBoundVar{E\sp{\prime}} \HOLBoundVar{X}) \HOLSymConst{\HOLTokenConj{}}
   (\HOLSymConst{\HOLTokenForall{}}\HOLBoundVar{E\sb{\mathrm{1}}} \HOLBoundVar{E\sb{\mathrm{2}}} \HOLBoundVar{E\sp{\prime}} \HOLBoundVar{X}.
      \HOLConst{CCS_Subst} (\HOLBoundVar{E\sb{\mathrm{1}}} \HOLSymConst{+} \HOLBoundVar{E\sb{\mathrm{2}}}) \HOLBoundVar{E\sp{\prime}} \HOLBoundVar{X} \HOLSymConst{=}
      \HOLConst{CCS_Subst} \HOLBoundVar{E\sb{\mathrm{1}}} \HOLBoundVar{E\sp{\prime}} \HOLBoundVar{X} \HOLSymConst{+} \HOLConst{CCS_Subst} \HOLBoundVar{E\sb{\mathrm{2}}} \HOLBoundVar{E\sp{\prime}} \HOLBoundVar{X}) \HOLSymConst{\HOLTokenConj{}}
   (\HOLSymConst{\HOLTokenForall{}}\HOLBoundVar{E\sb{\mathrm{1}}} \HOLBoundVar{E\sb{\mathrm{2}}} \HOLBoundVar{E\sp{\prime}} \HOLBoundVar{X}.
      \HOLConst{CCS_Subst} (\HOLBoundVar{E\sb{\mathrm{1}}} \HOLSymConst{||} \HOLBoundVar{E\sb{\mathrm{2}}}) \HOLBoundVar{E\sp{\prime}} \HOLBoundVar{X} \HOLSymConst{=}
      \HOLConst{CCS_Subst} \HOLBoundVar{E\sb{\mathrm{1}}} \HOLBoundVar{E\sp{\prime}} \HOLBoundVar{X} \HOLSymConst{||} \HOLConst{CCS_Subst} \HOLBoundVar{E\sb{\mathrm{2}}} \HOLBoundVar{E\sp{\prime}} \HOLBoundVar{X}) \HOLSymConst{\HOLTokenConj{}}
   (\HOLSymConst{\HOLTokenForall{}}\HOLBoundVar{L} \HOLBoundVar{E} \HOLBoundVar{E\sp{\prime}} \HOLBoundVar{X}.
      \HOLConst{CCS_Subst} (\HOLSymConst{\ensuremath{\nu}} \HOLBoundVar{L} \HOLBoundVar{E}) \HOLBoundVar{E\sp{\prime}} \HOLBoundVar{X} \HOLSymConst{=} \HOLSymConst{\ensuremath{\nu}} \HOLBoundVar{L} (\HOLConst{CCS_Subst} \HOLBoundVar{E} \HOLBoundVar{E\sp{\prime}} \HOLBoundVar{X})) \HOLSymConst{\HOLTokenConj{}}
   (\HOLSymConst{\HOLTokenForall{}}\HOLBoundVar{E} \HOLBoundVar{f} \HOLBoundVar{E\sp{\prime}} \HOLBoundVar{X}.
      \HOLConst{CCS_Subst} (\HOLConst{relab} \HOLBoundVar{E} \HOLBoundVar{f}) \HOLBoundVar{E\sp{\prime}} \HOLBoundVar{X} \HOLSymConst{=}
      \HOLConst{relab} (\HOLConst{CCS_Subst} \HOLBoundVar{E} \HOLBoundVar{E\sp{\prime}} \HOLBoundVar{X}) \HOLBoundVar{f}) \HOLSymConst{\HOLTokenConj{}}
   (\HOLSymConst{\HOLTokenForall{}}\HOLBoundVar{Y} \HOLBoundVar{E\sp{\prime}} \HOLBoundVar{X}.
      \HOLConst{CCS_Subst} (\HOLConst{var} \HOLBoundVar{Y}) \HOLBoundVar{E\sp{\prime}} \HOLBoundVar{X} \HOLSymConst{=} \HOLKeyword{if} \HOLBoundVar{Y} \HOLSymConst{=} \HOLBoundVar{X} \HOLKeyword{then} \HOLBoundVar{E\sp{\prime}} \HOLKeyword{else} \HOLConst{var} \HOLBoundVar{Y}) \HOLSymConst{\HOLTokenConj{}}
   \HOLSymConst{\HOLTokenForall{}}\HOLBoundVar{Y} \HOLBoundVar{E} \HOLBoundVar{E\sp{\prime}} \HOLBoundVar{X}.
     \HOLConst{CCS_Subst} (\HOLConst{rec} \HOLBoundVar{Y} \HOLBoundVar{E}) \HOLBoundVar{E\sp{\prime}} \HOLBoundVar{X} \HOLSymConst{=}
     \HOLKeyword{if} \HOLBoundVar{Y} \HOLSymConst{=} \HOLBoundVar{X} \HOLKeyword{then} \HOLConst{rec} \HOLBoundVar{Y} \HOLBoundVar{E} \HOLKeyword{else} \HOLConst{rec} \HOLBoundVar{Y} (\HOLConst{CCS_Subst} \HOLBoundVar{E} \HOLBoundVar{E\sp{\prime}} \HOLBoundVar{X})
\end{alltt}

In HOL4, any inductive relation defined by command \texttt{Hol_reln}
will return with three (well, actually four) theorems: 1) the
rules, 2) the induction (and strong induction) theorem and 3) the ``cases''
theorem. Only with all these theorems, the relation can be precisely
defined. For example, to prove certain CCS transitions are impossible,
the following long ``cases'' theorem (which asserts that the relation is a fixed
point) must be used:
\begin{alltt}
\HOLTokenTurnstile{} \HOLFreeVar{a\sb{\mathrm{0}}} --\HOLFreeVar{a\sb{\mathrm{1}}}-> \HOLFreeVar{a\sb{\mathrm{2}}} \HOLSymConst{\HOLTokenEquiv{}}
   (\HOLFreeVar{a\sb{\mathrm{0}}} \HOLSymConst{=} \HOLFreeVar{a\sb{\mathrm{1}}}\HOLSymConst{..}\HOLFreeVar{a\sb{\mathrm{2}}}) \HOLSymConst{\HOLTokenDisj{}} (\HOLSymConst{\HOLTokenExists{}}\HOLBoundVar{E} \HOLBoundVar{E\sp{\prime}}. (\HOLFreeVar{a\sb{\mathrm{0}}} \HOLSymConst{=} \HOLBoundVar{E} \HOLSymConst{+} \HOLBoundVar{E\sp{\prime}}) \HOLSymConst{\HOLTokenConj{}} \HOLBoundVar{E} --\HOLFreeVar{a\sb{\mathrm{1}}}-> \HOLFreeVar{a\sb{\mathrm{2}}}) \HOLSymConst{\HOLTokenDisj{}}
   (\HOLSymConst{\HOLTokenExists{}}\HOLBoundVar{E} \HOLBoundVar{E\sp{\prime}}. (\HOLFreeVar{a\sb{\mathrm{0}}} \HOLSymConst{=} \HOLBoundVar{E\sp{\prime}} \HOLSymConst{+} \HOLBoundVar{E}) \HOLSymConst{\HOLTokenConj{}} \HOLBoundVar{E} --\HOLFreeVar{a\sb{\mathrm{1}}}-> \HOLFreeVar{a\sb{\mathrm{2}}}) \HOLSymConst{\HOLTokenDisj{}}
   (\HOLSymConst{\HOLTokenExists{}}\HOLBoundVar{E} \HOLBoundVar{E\sb{\mathrm{1}}} \HOLBoundVar{E\sp{\prime}}. (\HOLFreeVar{a\sb{\mathrm{0}}} \HOLSymConst{=} \HOLBoundVar{E} \HOLSymConst{||} \HOLBoundVar{E\sp{\prime}}) \HOLSymConst{\HOLTokenConj{}} (\HOLFreeVar{a\sb{\mathrm{2}}} \HOLSymConst{=} \HOLBoundVar{E\sb{\mathrm{1}}} \HOLSymConst{||} \HOLBoundVar{E\sp{\prime}}) \HOLSymConst{\HOLTokenConj{}} \HOLBoundVar{E} --\HOLFreeVar{a\sb{\mathrm{1}}}-> \HOLBoundVar{E\sb{\mathrm{1}}}) \HOLSymConst{\HOLTokenDisj{}}
   (\HOLSymConst{\HOLTokenExists{}}\HOLBoundVar{E} \HOLBoundVar{E\sb{\mathrm{1}}} \HOLBoundVar{E\sp{\prime}}. (\HOLFreeVar{a\sb{\mathrm{0}}} \HOLSymConst{=} \HOLBoundVar{E\sp{\prime}} \HOLSymConst{||} \HOLBoundVar{E}) \HOLSymConst{\HOLTokenConj{}} (\HOLFreeVar{a\sb{\mathrm{2}}} \HOLSymConst{=} \HOLBoundVar{E\sp{\prime}} \HOLSymConst{||} \HOLBoundVar{E\sb{\mathrm{1}}}) \HOLSymConst{\HOLTokenConj{}} \HOLBoundVar{E} --\HOLFreeVar{a\sb{\mathrm{1}}}-> \HOLBoundVar{E\sb{\mathrm{1}}}) \HOLSymConst{\HOLTokenDisj{}}
   (\HOLSymConst{\HOLTokenExists{}}\HOLBoundVar{E} \HOLBoundVar{l} \HOLBoundVar{E\sb{\mathrm{1}}} \HOLBoundVar{E\sp{\prime}} \HOLBoundVar{E\sb{\mathrm{2}}}.
      (\HOLFreeVar{a\sb{\mathrm{0}}} \HOLSymConst{=} \HOLBoundVar{E} \HOLSymConst{||} \HOLBoundVar{E\sp{\prime}}) \HOLSymConst{\HOLTokenConj{}} (\HOLFreeVar{a\sb{\mathrm{1}}} \HOLSymConst{=} \HOLSymConst{\ensuremath{\tau}}) \HOLSymConst{\HOLTokenConj{}} (\HOLFreeVar{a\sb{\mathrm{2}}} \HOLSymConst{=} \HOLBoundVar{E\sb{\mathrm{1}}} \HOLSymConst{||} \HOLBoundVar{E\sb{\mathrm{2}}}) \HOLSymConst{\HOLTokenConj{}}
      \HOLBoundVar{E} --\HOLConst{label} \HOLBoundVar{l}-> \HOLBoundVar{E\sb{\mathrm{1}}} \HOLSymConst{\HOLTokenConj{}} \HOLBoundVar{E\sp{\prime}} --\HOLConst{label} (\HOLConst{COMPL} \HOLBoundVar{l})-> \HOLBoundVar{E\sb{\mathrm{2}}}) \HOLSymConst{\HOLTokenDisj{}}
   (\HOLSymConst{\HOLTokenExists{}}\HOLBoundVar{E} \HOLBoundVar{E\sp{\prime}} \HOLBoundVar{l} \HOLBoundVar{L}.
      (\HOLFreeVar{a\sb{\mathrm{0}}} \HOLSymConst{=} \HOLSymConst{\ensuremath{\nu}} \HOLBoundVar{L} \HOLBoundVar{E}) \HOLSymConst{\HOLTokenConj{}} (\HOLFreeVar{a\sb{\mathrm{2}}} \HOLSymConst{=} \HOLSymConst{\ensuremath{\nu}} \HOLBoundVar{L} \HOLBoundVar{E\sp{\prime}}) \HOLSymConst{\HOLTokenConj{}} \HOLBoundVar{E} --\HOLFreeVar{a\sb{\mathrm{1}}}-> \HOLBoundVar{E\sp{\prime}} \HOLSymConst{\HOLTokenConj{}}
      ((\HOLFreeVar{a\sb{\mathrm{1}}} \HOLSymConst{=} \HOLSymConst{\ensuremath{\tau}}) \HOLSymConst{\HOLTokenDisj{}} (\HOLFreeVar{a\sb{\mathrm{1}}} \HOLSymConst{=} \HOLConst{label} \HOLBoundVar{l}) \HOLSymConst{\HOLTokenConj{}} \HOLBoundVar{l} \HOLSymConst{\HOLTokenNotIn{}} \HOLBoundVar{L} \HOLSymConst{\HOLTokenConj{}} \HOLConst{COMPL} \HOLBoundVar{l} \HOLSymConst{\HOLTokenNotIn{}} \HOLBoundVar{L})) \HOLSymConst{\HOLTokenDisj{}}
   (\HOLSymConst{\HOLTokenExists{}}\HOLBoundVar{E} \HOLBoundVar{u} \HOLBoundVar{E\sp{\prime}} \HOLBoundVar{rf}.
      (\HOLFreeVar{a\sb{\mathrm{0}}} \HOLSymConst{=} \HOLConst{relab} \HOLBoundVar{E} \HOLBoundVar{rf}) \HOLSymConst{\HOLTokenConj{}} (\HOLFreeVar{a\sb{\mathrm{1}}} \HOLSymConst{=} \HOLConst{relabel} \HOLBoundVar{rf} \HOLBoundVar{u}) \HOLSymConst{\HOLTokenConj{}}
      (\HOLFreeVar{a\sb{\mathrm{2}}} \HOLSymConst{=} \HOLConst{relab} \HOLBoundVar{E\sp{\prime}} \HOLBoundVar{rf}) \HOLSymConst{\HOLTokenConj{}} \HOLBoundVar{E} --\HOLBoundVar{u}-> \HOLBoundVar{E\sp{\prime}}) \HOLSymConst{\HOLTokenDisj{}}
   \HOLSymConst{\HOLTokenExists{}}\HOLBoundVar{E} \HOLBoundVar{X}. (\HOLFreeVar{a\sb{\mathrm{0}}} \HOLSymConst{=} \HOLConst{rec} \HOLBoundVar{X} \HOLBoundVar{E}) \HOLSymConst{\HOLTokenConj{}} \HOLConst{CCS_Subst} \HOLBoundVar{E} (\HOLConst{rec} \HOLBoundVar{X} \HOLBoundVar{E}) \HOLBoundVar{X} --\HOLFreeVar{a\sb{\mathrm{1}}}-> \HOLFreeVar{a\sb{\mathrm{2}}}
\end{alltt}

Here are some results proved using above ``cases'' theorem (i. e. they cannot be
proved with only the SOS inference rules):
\begin{alltt}
NIL_NO_TRANS:
\HOLTokenTurnstile{} \HOLSymConst{\HOLTokenNeg{}}(\HOLConst{nil} --\HOLFreeVar{u}-> \HOLFreeVar{E})
TRANS_IMP_NO_NIL:
\HOLTokenTurnstile{} \HOLFreeVar{E} --\HOLFreeVar{u}-> \HOLFreeVar{E\sp{\prime}} \HOLSymConst{\HOLTokenImp{}} \HOLFreeVar{E} \HOLSymConst{\HOLTokenNotEqual{}} \HOLConst{nil}
TRANS_SUM_EQ:
\HOLTokenTurnstile{} \HOLFreeVar{E} \HOLSymConst{+} \HOLFreeVar{E\sp{\prime}} --\HOLFreeVar{u}-> \HOLFreeVar{E\sp{\prime\prime}} \HOLSymConst{\HOLTokenEquiv{}} \HOLFreeVar{E} --\HOLFreeVar{u}-> \HOLFreeVar{E\sp{\prime\prime}} \HOLSymConst{\HOLTokenDisj{}} \HOLFreeVar{E\sp{\prime}} --\HOLFreeVar{u}-> \HOLFreeVar{E\sp{\prime\prime}}
TRANS_PAR_EQ:
\HOLTokenTurnstile{} \HOLFreeVar{E} \HOLSymConst{||} \HOLFreeVar{E\sp{\prime}} --\HOLFreeVar{u}-> \HOLFreeVar{E\sp{\prime\prime}} \HOLSymConst{\HOLTokenEquiv{}}
   (\HOLSymConst{\HOLTokenExists{}}\HOLBoundVar{E\sb{\mathrm{1}}}. (\HOLFreeVar{E\sp{\prime\prime}} \HOLSymConst{=} \HOLBoundVar{E\sb{\mathrm{1}}} \HOLSymConst{||} \HOLFreeVar{E\sp{\prime}}) \HOLSymConst{\HOLTokenConj{}} \HOLFreeVar{E} --\HOLFreeVar{u}-> \HOLBoundVar{E\sb{\mathrm{1}}}) \HOLSymConst{\HOLTokenDisj{}}
   (\HOLSymConst{\HOLTokenExists{}}\HOLBoundVar{E\sb{\mathrm{1}}}. (\HOLFreeVar{E\sp{\prime\prime}} \HOLSymConst{=} \HOLFreeVar{E} \HOLSymConst{||} \HOLBoundVar{E\sb{\mathrm{1}}}) \HOLSymConst{\HOLTokenConj{}} \HOLFreeVar{E\sp{\prime}} --\HOLFreeVar{u}-> \HOLBoundVar{E\sb{\mathrm{1}}}) \HOLSymConst{\HOLTokenDisj{}}
   \HOLSymConst{\HOLTokenExists{}}\HOLBoundVar{E\sb{\mathrm{1}}} \HOLBoundVar{E\sb{\mathrm{2}}} \HOLBoundVar{l}.
     (\HOLFreeVar{u} \HOLSymConst{=} \HOLSymConst{\ensuremath{\tau}}) \HOLSymConst{\HOLTokenConj{}} (\HOLFreeVar{E\sp{\prime\prime}} \HOLSymConst{=} \HOLBoundVar{E\sb{\mathrm{1}}} \HOLSymConst{||} \HOLBoundVar{E\sb{\mathrm{2}}}) \HOLSymConst{\HOLTokenConj{}} \HOLFreeVar{E} --\HOLConst{label} \HOLBoundVar{l}-> \HOLBoundVar{E\sb{\mathrm{1}}} \HOLSymConst{\HOLTokenConj{}}
     \HOLFreeVar{E\sp{\prime}} --\HOLConst{label} (\HOLConst{COMPL} \HOLBoundVar{l})-> \HOLBoundVar{E\sb{\mathrm{2}}}
TRANS_RESTR_EQ:
\HOLTokenTurnstile{} \HOLSymConst{\ensuremath{\nu}} \HOLFreeVar{L} \HOLFreeVar{E} --\HOLFreeVar{u}-> \HOLFreeVar{E\sp{\prime}} \HOLSymConst{\HOLTokenEquiv{}}
   \HOLSymConst{\HOLTokenExists{}}\HOLBoundVar{E\sp{\prime\prime}} \HOLBoundVar{l}.
     (\HOLFreeVar{E\sp{\prime}} \HOLSymConst{=} \HOLSymConst{\ensuremath{\nu}} \HOLFreeVar{L} \HOLBoundVar{E\sp{\prime\prime}}) \HOLSymConst{\HOLTokenConj{}} \HOLFreeVar{E} --\HOLFreeVar{u}-> \HOLBoundVar{E\sp{\prime\prime}} \HOLSymConst{\HOLTokenConj{}}
     ((\HOLFreeVar{u} \HOLSymConst{=} \HOLSymConst{\ensuremath{\tau}}) \HOLSymConst{\HOLTokenDisj{}} (\HOLFreeVar{u} \HOLSymConst{=} \HOLConst{label} \HOLBoundVar{l}) \HOLSymConst{\HOLTokenConj{}} \HOLBoundVar{l} \HOLSymConst{\HOLTokenNotIn{}} \HOLFreeVar{L} \HOLSymConst{\HOLTokenConj{}} \HOLConst{COMPL} \HOLBoundVar{l} \HOLSymConst{\HOLTokenNotIn{}} \HOLFreeVar{L})
\end{alltt}

Finally, it's worth to mention that, the following induction theorem
generated by \texttt{Hol_reln} was never used (nor needed) in this
project:
\begin{alltt}
TRANS_ind:
\HOLTokenTurnstile{} (\HOLSymConst{\HOLTokenForall{}}\HOLBoundVar{E} \HOLBoundVar{u}. \HOLFreeVar{TRANS\sp{\prime}} (\HOLBoundVar{u}\HOLSymConst{..}\HOLBoundVar{E}) \HOLBoundVar{u} \HOLBoundVar{E}) \HOLSymConst{\HOLTokenConj{}}
   (\HOLSymConst{\HOLTokenForall{}}\HOLBoundVar{E} \HOLBoundVar{u} \HOLBoundVar{E\sb{\mathrm{1}}} \HOLBoundVar{E\sp{\prime}}. \HOLFreeVar{TRANS\sp{\prime}} \HOLBoundVar{E} \HOLBoundVar{u} \HOLBoundVar{E\sb{\mathrm{1}}} \HOLSymConst{\HOLTokenImp{}} \HOLFreeVar{TRANS\sp{\prime}} (\HOLBoundVar{E} \HOLSymConst{+} \HOLBoundVar{E\sp{\prime}}) \HOLBoundVar{u} \HOLBoundVar{E\sb{\mathrm{1}}}) \HOLSymConst{\HOLTokenConj{}}
   (\HOLSymConst{\HOLTokenForall{}}\HOLBoundVar{E} \HOLBoundVar{u} \HOLBoundVar{E\sb{\mathrm{1}}} \HOLBoundVar{E\sp{\prime}}. \HOLFreeVar{TRANS\sp{\prime}} \HOLBoundVar{E} \HOLBoundVar{u} \HOLBoundVar{E\sb{\mathrm{1}}} \HOLSymConst{\HOLTokenImp{}} \HOLFreeVar{TRANS\sp{\prime}} (\HOLBoundVar{E\sp{\prime}} \HOLSymConst{+} \HOLBoundVar{E}) \HOLBoundVar{u} \HOLBoundVar{E\sb{\mathrm{1}}}) \HOLSymConst{\HOLTokenConj{}}
   (\HOLSymConst{\HOLTokenForall{}}\HOLBoundVar{E} \HOLBoundVar{u} \HOLBoundVar{E\sb{\mathrm{1}}} \HOLBoundVar{E\sp{\prime}}.
      \HOLFreeVar{TRANS\sp{\prime}} \HOLBoundVar{E} \HOLBoundVar{u} \HOLBoundVar{E\sb{\mathrm{1}}} \HOLSymConst{\HOLTokenImp{}} \HOLFreeVar{TRANS\sp{\prime}} (\HOLBoundVar{E} \HOLSymConst{||} \HOLBoundVar{E\sp{\prime}}) \HOLBoundVar{u} (\HOLBoundVar{E\sb{\mathrm{1}}} \HOLSymConst{||} \HOLBoundVar{E\sp{\prime}})) \HOLSymConst{\HOLTokenConj{}}
   (\HOLSymConst{\HOLTokenForall{}}\HOLBoundVar{E} \HOLBoundVar{u} \HOLBoundVar{E\sb{\mathrm{1}}} \HOLBoundVar{E\sp{\prime}}.
      \HOLFreeVar{TRANS\sp{\prime}} \HOLBoundVar{E} \HOLBoundVar{u} \HOLBoundVar{E\sb{\mathrm{1}}} \HOLSymConst{\HOLTokenImp{}} \HOLFreeVar{TRANS\sp{\prime}} (\HOLBoundVar{E\sp{\prime}} \HOLSymConst{||} \HOLBoundVar{E}) \HOLBoundVar{u} (\HOLBoundVar{E\sp{\prime}} \HOLSymConst{||} \HOLBoundVar{E\sb{\mathrm{1}}})) \HOLSymConst{\HOLTokenConj{}}
   (\HOLSymConst{\HOLTokenForall{}}\HOLBoundVar{E} \HOLBoundVar{l} \HOLBoundVar{E\sb{\mathrm{1}}} \HOLBoundVar{E\sp{\prime}} \HOLBoundVar{E\sb{\mathrm{2}}}.
      \HOLFreeVar{TRANS\sp{\prime}} \HOLBoundVar{E} (\HOLConst{label} \HOLBoundVar{l}) \HOLBoundVar{E\sb{\mathrm{1}}} \HOLSymConst{\HOLTokenConj{}} \HOLFreeVar{TRANS\sp{\prime}} \HOLBoundVar{E\sp{\prime}} (\HOLConst{label} (\HOLConst{COMPL} \HOLBoundVar{l})) \HOLBoundVar{E\sb{\mathrm{2}}} \HOLSymConst{\HOLTokenImp{}}
      \HOLFreeVar{TRANS\sp{\prime}} (\HOLBoundVar{E} \HOLSymConst{||} \HOLBoundVar{E\sp{\prime}}) \HOLSymConst{\ensuremath{\tau}} (\HOLBoundVar{E\sb{\mathrm{1}}} \HOLSymConst{||} \HOLBoundVar{E\sb{\mathrm{2}}})) \HOLSymConst{\HOLTokenConj{}}
   (\HOLSymConst{\HOLTokenForall{}}\HOLBoundVar{E} \HOLBoundVar{u} \HOLBoundVar{E\sp{\prime}} \HOLBoundVar{l} \HOLBoundVar{L}.
      \HOLFreeVar{TRANS\sp{\prime}} \HOLBoundVar{E} \HOLBoundVar{u} \HOLBoundVar{E\sp{\prime}} \HOLSymConst{\HOLTokenConj{}}
      ((\HOLBoundVar{u} \HOLSymConst{=} \HOLSymConst{\ensuremath{\tau}}) \HOLSymConst{\HOLTokenDisj{}} (\HOLBoundVar{u} \HOLSymConst{=} \HOLConst{label} \HOLBoundVar{l}) \HOLSymConst{\HOLTokenConj{}} \HOLBoundVar{l} \HOLSymConst{\HOLTokenNotIn{}} \HOLBoundVar{L} \HOLSymConst{\HOLTokenConj{}} \HOLConst{COMPL} \HOLBoundVar{l} \HOLSymConst{\HOLTokenNotIn{}} \HOLBoundVar{L}) \HOLSymConst{\HOLTokenImp{}}
      \HOLFreeVar{TRANS\sp{\prime}} (\HOLSymConst{\ensuremath{\nu}} \HOLBoundVar{L} \HOLBoundVar{E}) \HOLBoundVar{u} (\HOLSymConst{\ensuremath{\nu}} \HOLBoundVar{L} \HOLBoundVar{E\sp{\prime}})) \HOLSymConst{\HOLTokenConj{}}
   (\HOLSymConst{\HOLTokenForall{}}\HOLBoundVar{E} \HOLBoundVar{u} \HOLBoundVar{E\sp{\prime}} \HOLBoundVar{rf}.
      \HOLFreeVar{TRANS\sp{\prime}} \HOLBoundVar{E} \HOLBoundVar{u} \HOLBoundVar{E\sp{\prime}} \HOLSymConst{\HOLTokenImp{}}
      \HOLFreeVar{TRANS\sp{\prime}} (\HOLConst{relab} \HOLBoundVar{E} \HOLBoundVar{rf}) (\HOLConst{relabel} \HOLBoundVar{rf} \HOLBoundVar{u}) (\HOLConst{relab} \HOLBoundVar{E\sp{\prime}} \HOLBoundVar{rf})) \HOLSymConst{\HOLTokenConj{}}
   (\HOLSymConst{\HOLTokenForall{}}\HOLBoundVar{E} \HOLBoundVar{u} \HOLBoundVar{X} \HOLBoundVar{E\sb{\mathrm{1}}}.
      \HOLFreeVar{TRANS\sp{\prime}} (\HOLConst{CCS_Subst} \HOLBoundVar{E} (\HOLConst{rec} \HOLBoundVar{X} \HOLBoundVar{E}) \HOLBoundVar{X}) \HOLBoundVar{u} \HOLBoundVar{E\sb{\mathrm{1}}} \HOLSymConst{\HOLTokenImp{}}
      \HOLFreeVar{TRANS\sp{\prime}} (\HOLConst{rec} \HOLBoundVar{X} \HOLBoundVar{E}) \HOLBoundVar{u} \HOLBoundVar{E\sb{\mathrm{1}}}) \HOLSymConst{\HOLTokenImp{}}
   \HOLSymConst{\HOLTokenForall{}}\HOLBoundVar{a\sb{\mathrm{0}}} \HOLBoundVar{a\sb{\mathrm{1}}} \HOLBoundVar{a\sb{\mathrm{2}}}. \HOLBoundVar{a\sb{\mathrm{0}}} --\HOLBoundVar{a\sb{\mathrm{1}}}-> \HOLBoundVar{a\sb{\mathrm{2}}} \HOLSymConst{\HOLTokenImp{}} \HOLFreeVar{TRANS\sp{\prime}} \HOLBoundVar{a\sb{\mathrm{0}}} \HOLBoundVar{a\sb{\mathrm{1}}} \HOLBoundVar{a\sb{\mathrm{2}}}
\end{alltt}
The purpose of above induction theorem is to assert the transition
relation to the least fixed point of the function generated from SOS
inference rules. On the other side, if we define a co-inductive
relation from the same SOS rules, we get the same rules and ``cases''
theorems, and the only difference is another co-induction theorem in
place of above induction theorem. This seems indicating that, the
least fixed point coincides with greatest fixed point for Finitary
CCS . This result is never formally proved, but both Prof. Gorrieri and
the author believe it's true. (However, Prof. Gorrieri thinks it's
NOT appropriate to use co-induction in this case, unless infinite sums and parallels were defined as part of CCS syntax)

Also, we have noticed that, to prove certain CCS transition is
impossible, it's enough to use just the above ``cases'' theorem. Since
the CCS datatype itself is inductively defined, therefore already
Finitary,  all those invalid transitions seems must be outside of the
fixed point, in another word, they're even outside of the greatest
fixed point. If one day we had changed the definition of CCS datatype to
allow infinite sums and parallels, the SOS inference rules should
still work, but the transition relation should be then defined
co-inductively, to allow valid transitions for both finitary and
infinitary CCS processes.\footnote{Michael Norrish has different opinion with the following
argument: ``Infinite sums and parallels would not require a
coinductive definition.  Coinductive definitions give you “infinite
depth”.  Infinite sums and parallels would only require infinite
breadth.'' However, a CCS transition which is inside the fixed point
but outside of the least fixed point, is yet to be found to support
this argument.}

\subsection{Decision procedure for CCS transitions}

It's possible to use SOS inference rules and theorems derived from
them for proving theorems about the transitions between any two CCS
processes. However, what's more useful is the decision procedure
which automatically decide all possible transitions and formally prove them.

For any CCS process, there is a decision procedure as a recursive
function, which can completely decide all its possible (one-step)
transitions. In HOL, this decision procedure can be implemented as a
normal Standard ML function \texttt{CCS\_TRANS\_CONV} of type
\texttt{term -> theorem}, the returned theorem fully characterize the
possible transitions of the input CCS process.

For instance, we know that the process $(a.0 | \bar{a}.0)$ have three
possible transitions:
\begin{enumerate}
\item $(a.0 | \bar{a}.0) \overset{a}{\longrightarrow} (0 | \bar{a}.0)$;
\item $(a.0 | \bar{a}.0) \overset{\bar{a}}{\longrightarrow} (a.0 | 0)$;
\item $(a.0 | \bar{a}.0) \overset{\tau}{\longrightarrow} (0 | 0)$.
\end{enumerate}
To completely decide all possible transitions, if done manually, the following work should be done:
\begin{enumerate}
\item Prove there exists transitions from $(a.0 | \bar{a}.0)$ (optionally);
\item Prove each of above three transitions using SOS inference rules;
\item Prove there's no other transitions, using the ``cases'' theorems
  generated from the \texttt{TRANS} relation.
\end{enumerate}
Here are the related theorems manually proved:
\begin{alltt}
r1_has_trans:
\HOLTokenTurnstile{} \HOLSymConst{\HOLTokenExists{}}\HOLBoundVar{l} \HOLBoundVar{G}. \HOLConst{In} \HOLStringLit{a}\HOLSymConst{..}\HOLConst{nil} \HOLSymConst{||} \HOLConst{Out} \HOLStringLit{a}\HOLSymConst{..}\HOLConst{nil} --\HOLBoundVar{l}-> \HOLBoundVar{G}
r1_trans_1:
\HOLTokenTurnstile{} \HOLConst{In} \HOLStringLit{a}\HOLSymConst{..}\HOLConst{nil} \HOLSymConst{||} \HOLConst{Out} \HOLStringLit{a}\HOLSymConst{..}\HOLConst{nil} --\HOLConst{In} \HOLStringLit{a}-> \HOLConst{nil} \HOLSymConst{||} \HOLConst{Out} \HOLStringLit{a}\HOLSymConst{..}\HOLConst{nil}
r1_trans_2:
\HOLTokenTurnstile{} \HOLConst{In} \HOLStringLit{a}\HOLSymConst{..}\HOLConst{nil} \HOLSymConst{||} \HOLConst{Out} \HOLStringLit{a}\HOLSymConst{..}\HOLConst{nil} --\HOLConst{Out} \HOLStringLit{a}-> \HOLConst{In} \HOLStringLit{a}\HOLSymConst{..}\HOLConst{nil} \HOLSymConst{||} \HOLConst{nil}
r1_trans_3:
\HOLTokenTurnstile{} \HOLConst{In} \HOLStringLit{a}\HOLSymConst{..}\HOLConst{nil} \HOLSymConst{||} \HOLConst{Out} \HOLStringLit{a}\HOLSymConst{..}\HOLConst{nil} --\HOLSymConst{\ensuremath{\tau}}-> \HOLConst{nil} \HOLSymConst{||} \HOLConst{nil}
r1_has_no_other_trans:
\HOLTokenTurnstile{} \HOLSymConst{\HOLTokenNeg{}}\HOLSymConst{\HOLTokenExists{}}\HOLBoundVar{l} \HOLBoundVar{G}.
      \HOLSymConst{\HOLTokenNeg{}}((\HOLBoundVar{G} \HOLSymConst{=} \HOLConst{nil} \HOLSymConst{||} \HOLConst{Out} \HOLStringLit{a}\HOLSymConst{..}\HOLConst{nil}) \HOLSymConst{\HOLTokenConj{}} (\HOLBoundVar{l} \HOLSymConst{=} \HOLConst{In} \HOLStringLit{a}) \HOLSymConst{\HOLTokenDisj{}}
        (\HOLBoundVar{G} \HOLSymConst{=} \HOLConst{In} \HOLStringLit{a}\HOLSymConst{..}\HOLConst{nil} \HOLSymConst{||} \HOLConst{nil}) \HOLSymConst{\HOLTokenConj{}} (\HOLBoundVar{l} \HOLSymConst{=} \HOLConst{Out} \HOLStringLit{a}) \HOLSymConst{\HOLTokenDisj{}}
        (\HOLBoundVar{G} \HOLSymConst{=} \HOLConst{nil} \HOLSymConst{||} \HOLConst{nil}) \HOLSymConst{\HOLTokenConj{}} (\HOLBoundVar{l} \HOLSymConst{=} \HOLSymConst{\ensuremath{\tau}})) \HOLSymConst{\HOLTokenConj{}}
      \HOLConst{In} \HOLStringLit{a}\HOLSymConst{..}\HOLConst{nil} \HOLSymConst{||} \HOLConst{Out} \HOLStringLit{a}\HOLSymConst{..}\HOLConst{nil} --\HOLBoundVar{l}-> \HOLBoundVar{G}
\end{alltt}

Instead, if we use the function \texttt{CCS\_TRANS\_CONV} with the
root process:
\begin{lstlisting}
> CCS_TRANS_CONV
	 ``par (prefix (label (name "a")) nil)
	       (prefix (label (coname "a")) nil)``
\end{lstlisting}
As the result, the following theorem is returned:
\begin{alltt}
ex_A: \HOLTokenTurnstile{} \HOLConst{In} \HOLStringLit{a}\HOLSymConst{..}\HOLConst{nil} \HOLSymConst{||} \HOLConst{Out} \HOLStringLit{a}\HOLSymConst{..}\HOLConst{nil} --\HOLFreeVar{u}-> \HOLFreeVar{E} \HOLSymConst{\HOLTokenEquiv{}}
   ((\HOLFreeVar{u} \HOLSymConst{=} \HOLConst{In} \HOLStringLit{a}) \HOLSymConst{\HOLTokenConj{}} (\HOLFreeVar{E} \HOLSymConst{=} \HOLConst{nil} \HOLSymConst{||} \HOLConst{Out} \HOLStringLit{a}\HOLSymConst{..}\HOLConst{nil}) \HOLSymConst{\HOLTokenDisj{}}
    (\HOLFreeVar{u} \HOLSymConst{=} \HOLConst{Out} \HOLStringLit{a}) \HOLSymConst{\HOLTokenConj{}} (\HOLFreeVar{E} \HOLSymConst{=} \HOLConst{In} \HOLStringLit{a}\HOLSymConst{..}\HOLConst{nil} \HOLSymConst{||} \HOLConst{nil})) \HOLSymConst{\HOLTokenDisj{}}
   (\HOLFreeVar{u} \HOLSymConst{=} \HOLSymConst{\ensuremath{\tau}}) \HOLSymConst{\HOLTokenConj{}} (\HOLFreeVar{E} \HOLSymConst{=} \HOLConst{nil} \HOLSymConst{||} \HOLConst{nil})
\end{alltt}
From this theorem, we can see there're only three possible transitions
and there's no others. Therefore it contains all information expressed
by previous manually proved 5 theorems (in theory we can also try to
manually prove this single theorem, but it's not easy since the steps
required will be at least the sum of all previous proofs).

As a further example, if we put a restriction on label ``a'' and check
the process $(\nu a)(a.0 |
\bar{a}.0)$ instead, there will be only one possible transition:
\begin{alltt}
ex_B: \HOLTokenTurnstile{} \HOLSymConst{\ensuremath{\nu}} \HOLStringLit{a} (\HOLConst{In} \HOLStringLit{a}\HOLSymConst{..}\HOLConst{nil} \HOLSymConst{||} \HOLConst{Out} \HOLStringLit{a}\HOLSymConst{..}\HOLConst{nil}) --\HOLFreeVar{u}-> \HOLFreeVar{E} \HOLSymConst{\HOLTokenEquiv{}}
   (\HOLFreeVar{u} \HOLSymConst{=} \HOLSymConst{\ensuremath{\tau}}) \HOLSymConst{\HOLTokenConj{}} (\HOLFreeVar{E} \HOLSymConst{=} \HOLSymConst{\ensuremath{\nu}} \HOLStringLit{a} (\HOLConst{nil} \HOLSymConst{||} \HOLConst{nil}))
\end{alltt}

It's possible to extract a list of possible transitions together with
the actions, into a list. This work can be done automatically by the
function \texttt{strip_trans}. Finally, if both the theorem and the
list of transitions are needed, the function \texttt{CCS\_TRANS} and
its compact-form variant \texttt{CCS\_TRANS'} can be used. For the
previous example process $(a.0 | \bar{a}.0)$, calling
\texttt{CCS\_TRANS'} on it in HOL's interactive environment has the
following results:
\begin{lstlisting}
> CCS_TRANS ``In "a"..nil || Out "a"..nil``;
val it =
   (|- !u E.
     In "a"..nil || Out "a"..nil --u-> E <=>
     ((u = In "a") /\ (E = nil || Out "a"..nil) \/
      (u = Out "a") /\ (E = In "a"..nil || nil)) \/
     (u = tau) /\ (E = nil || nil),
    [(``In "a"``,
      ``nil || Out "a"..nil``),
     (``Out "a"``,
      ``In "a"..nil || nil``),
     (``'t``,
      ``nil || nil``)]):
   thm * (term * term) list
\end{lstlisting}

The main function \texttt{CCS_TRANS_CONV} is implemented in about 500
lines of Standard ML code, and it depends on many customized tacticals,
and functions to access the internal structure of CCS related theorem
and terms.  We have tried our best to make sure the correctness of
this function, but certain bugs are still inevitable.\footnote{If the
  internal proof constructed in the function is wrong, then the function won't
  return a theorem. But if the function successfully returns a theorem,
  the proof for this theorem must be correct, because there's no other
  way to return a theorem except for correctly proving it in HOL
  theorem prover.} However, since
it's implemented in theorem prover, and the return value of this
function is a theorem, what we can guarentee is the following fact:
\begin{quote}
Whenever the function terminates with a theorem returned, as long as
the theorem has ``correct'' forms, the CCS
transitions indicated in the returned theorem is indeed all possible
transitions from the input process. No matter if there're bugs in our program.
\end{quote}

In another words, any remain bug in the program can only stop the
whole function for returning a result, but as long as the result is
returned, it cannot be wrong (i.e. a fake theorem). This sounds like a
different kind of trusted computing than normal sense. In general, for
any algorithm implemented in any normal programming langauges, since
the output is just a primitive value or data structure which can be
arbitrary constructed or changed due to potential bugs in the
implementation, the only way to trust these results, is to have the
entire program carefully modelled and verified. But in our case, the
Standard ML program code is not verified, but the result (once appears) can still be
fully trusted, isn't this amazing?

\subsection{Strong bisimulation, strong equivalence and co-induction}

The concept of \emph{Bisimulation} (and \emph{Bisimulation
  Equivalence} with variants) stands at the
central position of Concurrency Theory, as one major approach of model
checking is to check the bisimulation equivalence between the
specification and implementation of the same reactive system.
Besides, it's well known that,
Strong Equivalence as a relation, must be defined
\emph{co-inductively}. (And in fact, strong equivalence is one of the
most well-studied co-inductive relation in computer
science. \cite{Sangiorgi:2011ut}) In this section, we study the
definition of strong and weak
bisimulation and (bisimulation) equivalences, and their
possible formalizations in HOL.

Recall the standard definition of strong bisimulation and strong equivalence
(c.f. p.43 of \cite{Gorrieri:2015jt}):
\begin{definition}{((Strong) bisimulation and (strong) bisimulation equivalence)}
Let $TS = (Q, A, \rightarrow)$ be a transition system. A
\emph{bisimulation} is a relation $R \subset Q \times Q$ such that $R$
and its inverse $R^{-1}$ are both simulation relations. More
explicitly, a bisimulation is a relation $R$ such that if
$(q_1,q_2)\in R$ then for all $\mu\in A$
\begin{itemize}
\item $\forall q_1' \text{ such that } q_1
  \overset{\mu}{\longrightarrow} q_1', \exists q_2' \text{ such that }
  q_2 \overset{\mu}{\longrightarrow} q_2' \text{ and } (q_1', q_2')
  \in R$,
\item $\forall q_2' \text{ such that } q_2
  \overset{\mu}{\longrightarrow} q_2', \exists q_1' \text{ such that }
  q_1 \overset{\mu}{\longrightarrow} q_1' \text{ and } (q_1', q_2')
  \in R$.
\end{itemize}
Two states $q$ and $q'$ are \emph{bisimular} (or \emph{bisimulation
  equivalent}), denoted $q \sim q'$, if there exists a bisimulation
$R$ such that $(q, q') \in R$.
\end{definition}
Noticed that, although above definition is expressed in LTS, it's also
applicable to CCS in which each process has the semantic model as a
rooted LTS. Given the fact that, all states involved in above
definition are target states of direct or indirect transition of the
initial pair of states, above definition can be directly used for CCS.

In HOL88, there's no way to define co-inductive relation
directly. However, it's possible to follow above definition literallly
and define bisimulation first, then define the bisimulation
equivalence on top of bisimulation. Here are the definitions 
translated from HOL88 to HOL4:
\begin{alltt}
\HOLTokenTurnstile{} \HOLConst{STRONG_BISIM} \HOLFreeVar{Bsm} \HOLSymConst{\HOLTokenEquiv{}}
   \HOLSymConst{\HOLTokenForall{}}\HOLBoundVar{E} \HOLBoundVar{E\sp{\prime}}.
     \HOLFreeVar{Bsm} \HOLBoundVar{E} \HOLBoundVar{E\sp{\prime}} \HOLSymConst{\HOLTokenImp{}}
     \HOLSymConst{\HOLTokenForall{}}\HOLBoundVar{u}.
       (\HOLSymConst{\HOLTokenForall{}}\HOLBoundVar{E\sb{\mathrm{1}}}. \HOLBoundVar{E} --\HOLBoundVar{u}-> \HOLBoundVar{E\sb{\mathrm{1}}} \HOLSymConst{\HOLTokenImp{}} \HOLSymConst{\HOLTokenExists{}}\HOLBoundVar{E\sb{\mathrm{2}}}. \HOLBoundVar{E\sp{\prime}} --\HOLBoundVar{u}-> \HOLBoundVar{E\sb{\mathrm{2}}} \HOLSymConst{\HOLTokenConj{}} \HOLFreeVar{Bsm} \HOLBoundVar{E\sb{\mathrm{1}}} \HOLBoundVar{E\sb{\mathrm{2}}}) \HOLSymConst{\HOLTokenConj{}}
       \HOLSymConst{\HOLTokenForall{}}\HOLBoundVar{E\sb{\mathrm{2}}}. \HOLBoundVar{E\sp{\prime}} --\HOLBoundVar{u}-> \HOLBoundVar{E\sb{\mathrm{2}}} \HOLSymConst{\HOLTokenImp{}} \HOLSymConst{\HOLTokenExists{}}\HOLBoundVar{E\sb{\mathrm{1}}}. \HOLBoundVar{E} --\HOLBoundVar{u}-> \HOLBoundVar{E\sb{\mathrm{1}}} \HOLSymConst{\HOLTokenConj{}} \HOLFreeVar{Bsm} \HOLBoundVar{E\sb{\mathrm{1}}} \HOLBoundVar{E\sb{\mathrm{2}}}
\HOLTokenTurnstile{} \HOLFreeVar{E} \HOLSymConst{\HOLTokenStrongEquiv} \HOLFreeVar{E\sp{\prime}} \HOLSymConst{\HOLTokenEquiv{}} \HOLSymConst{\HOLTokenExists{}}\HOLBoundVar{Bsm}. \HOLBoundVar{Bsm} \HOLFreeVar{E} \HOLFreeVar{E\sp{\prime}} \HOLSymConst{\HOLTokenConj{}} \HOLConst{STRONG_BISIM} \HOLBoundVar{Bsm}
\end{alltt}
From the second definition, we can see that, $q \sim q'$ if there
exists a bisimulation containing the pair $(q, q')$. This means that
$\sim$ is the union of all bisimulations, i.e.,
\begin{equation*}
\sim = \bigcup \{ R \subset Q \times Q \colon R \text{ is a
  bisimulation} \}.
\end{equation*}
In HOL4, the last formula can be proved with the notion ``bigunion'' in HOL's
\texttt{pred_setTheory} used. The only thing needed from CCS is above
deifnition of strong equivalence (the definition of strong
bisimulation is not needed at all):
\begin{alltt}
\HOLTokenTurnstile{} \HOLConst{STRONG_EQUIV} \HOLSymConst{=} \HOLConst{CURRY} (\HOLConst{BIGUNION} \HOLTokenLeftbrace{}\HOLConst{UNCURRY} \HOLBoundVar{R} \HOLTokenBar{} \HOLConst{STRONG_BISIM} \HOLBoundVar{R}\HOLTokenRightbrace{})
\end{alltt}
However, this theorem is not very useful for proving other
results. And the use of \texttt{CURRY} and \texttt{UNCURRY} is to
transform the relation from types between \HOLinline{\HOLTyOp{CCS} -> \HOLTyOp{CCS} -> \HOLTyOp{bool}}
and \HOLinline{\HOLTyOp{CCS} \HOLTokenProd{} \HOLTyOp{CCS} -> \HOLTyOp{bool}}, since relations in HOL cannot be
treated directly as mathematical sets.

The other way to define strong bisimulation equivalence is through
the fixed point of the following function $F$: (c.f. p.72 of \cite{Gorrieri:2015jt})
\begin{definition}
Given an LTS $(Q, A, \rightarrow)$, the function $F\colon \wp(Q\times
Q) \rightarrow \wp(Q\times Q)$ (i.e., a transformer of binary
relations over $Q$) is defined as follows. If $R \subset Q\times Q$,
then $(q_1,q_2) \in F(R)$ if and only if for all $\mu \in A$
\begin{itemize}
\item $\forall q_1' \text{ such that } q_1 \overset{\mu}{\longrightarrow} q_1', \exists
  q_2' \text{ such that } q_2 \overset{\mu}{\longrightarrow} q_2'
  \text{ and } (q_1', q_2') \in R$,
\item $\forall q_2' \text{ such that } q_2 \overset{\mu}{\longrightarrow} q_2', \exists
  q_1' \text{ such that } q_1 \overset{\mu}{\longrightarrow} q_1'
  \text{ and } (q_1', q_2') \in R$.
\end{itemize}
\end{definition}
And we can see by comparing the definition of above function and the
definition of bisimulation that (no formal proofs):
\begin{enumerate}
\item The function $F$ is monotone, i.e. if $R_1 \subset R_2$ then
  $F(R_1) \subset F(R_2)$.
\item A relation $R\subset Q\times Q$ is a bisimulation if and only if
  $R \subset F(R)$.
\end{enumerate}
Then according to Knaster-Tarski Fixed Point theorem, strong
bisimilarity $\sim$ is the greatest fixed point of $F$. And this is
also the definition of co-inductive relation defined by the same rules.

In HOL4, since the release Kananaskis-11, there's a new facility for
defining co-inductive relation. The entry command is
\texttt{Hol_coreln}, which has the same syntax as \texttt{Hol_reln}
for definining inductive relations. Using \texttt{Hol_coreln}, it's
possible to define the bisimulation equivalence \emph{directly} in
this way: (here we has chosen a new relation name
\texttt{STRONG_EQ})\footnote{Whenever ASCII-based HOL proof scripts were directly
  pasted, please understand the letter ``\texttt{!}'' as $\forall$,
  and ``\texttt{?}'' as $\exists$. They're part of HOL's term syntax. \cite{Anonymous:Iu-sOoz1}}
\begin{lstlisting}
val (STRONG_EQ_rules, STRONG_EQ_coind, STRONG_EQ_cases) = Hol_coreln `
    (!E E'.
       (!u.
         (!E1. TRANS E u E1 ==> 
               (?E2. TRANS E' u E2 /\ STRONG_EQ E1 E2)) /\
         (!E2. TRANS E' u E2 ==> 
               (?E1. TRANS E u E1 /\ STRONG_EQ E1 E2))) ==> STRONG_EQ E E')`;
\end{lstlisting}
HOL automatically generated 3 theorems from above definition:
\begin{alltt}
STRONG_EQ_rules:
\HOLTokenTurnstile{} (\HOLSymConst{\HOLTokenForall{}}\HOLBoundVar{u}.
      (\HOLSymConst{\HOLTokenForall{}}\HOLBoundVar{E\sb{\mathrm{1}}}. \HOLFreeVar{E} --\HOLBoundVar{u}-> \HOLBoundVar{E\sb{\mathrm{1}}} \HOLSymConst{\HOLTokenImp{}} \HOLSymConst{\HOLTokenExists{}}\HOLBoundVar{E\sb{\mathrm{2}}}. \HOLFreeVar{E\sp{\prime}} --\HOLBoundVar{u}-> \HOLBoundVar{E\sb{\mathrm{2}}} \HOLSymConst{\HOLTokenConj{}} \HOLConst{STRONG_EQ} \HOLBoundVar{E\sb{\mathrm{1}}} \HOLBoundVar{E\sb{\mathrm{2}}}) \HOLSymConst{\HOLTokenConj{}}
      \HOLSymConst{\HOLTokenForall{}}\HOLBoundVar{E\sb{\mathrm{2}}}. \HOLFreeVar{E\sp{\prime}} --\HOLBoundVar{u}-> \HOLBoundVar{E\sb{\mathrm{2}}} \HOLSymConst{\HOLTokenImp{}} \HOLSymConst{\HOLTokenExists{}}\HOLBoundVar{E\sb{\mathrm{1}}}. \HOLFreeVar{E} --\HOLBoundVar{u}-> \HOLBoundVar{E\sb{\mathrm{1}}} \HOLSymConst{\HOLTokenConj{}} \HOLConst{STRONG_EQ} \HOLBoundVar{E\sb{\mathrm{1}}} \HOLBoundVar{E\sb{\mathrm{2}}}) \HOLSymConst{\HOLTokenImp{}}
   \HOLConst{STRONG_EQ} \HOLFreeVar{E} \HOLFreeVar{E\sp{\prime}}
STRONG_EQ_coind:
\HOLTokenTurnstile{} (\HOLSymConst{\HOLTokenForall{}}\HOLBoundVar{a\sb{\mathrm{0}}} \HOLBoundVar{a\sb{\mathrm{1}}}.
      \HOLFreeVar{STRONG\HOLTokenUnderscore{}EQ\sp{\prime}} \HOLBoundVar{a\sb{\mathrm{0}}} \HOLBoundVar{a\sb{\mathrm{1}}} \HOLSymConst{\HOLTokenImp{}}
      \HOLSymConst{\HOLTokenForall{}}\HOLBoundVar{u}.
        (\HOLSymConst{\HOLTokenForall{}}\HOLBoundVar{E\sb{\mathrm{1}}}.
           \HOLBoundVar{a\sb{\mathrm{0}}} --\HOLBoundVar{u}-> \HOLBoundVar{E\sb{\mathrm{1}}} \HOLSymConst{\HOLTokenImp{}} \HOLSymConst{\HOLTokenExists{}}\HOLBoundVar{E\sb{\mathrm{2}}}. \HOLBoundVar{a\sb{\mathrm{1}}} --\HOLBoundVar{u}-> \HOLBoundVar{E\sb{\mathrm{2}}} \HOLSymConst{\HOLTokenConj{}} \HOLFreeVar{STRONG\HOLTokenUnderscore{}EQ\sp{\prime}} \HOLBoundVar{E\sb{\mathrm{1}}} \HOLBoundVar{E\sb{\mathrm{2}}}) \HOLSymConst{\HOLTokenConj{}}
        \HOLSymConst{\HOLTokenForall{}}\HOLBoundVar{E\sb{\mathrm{2}}}.
          \HOLBoundVar{a\sb{\mathrm{1}}} --\HOLBoundVar{u}-> \HOLBoundVar{E\sb{\mathrm{2}}} \HOLSymConst{\HOLTokenImp{}} \HOLSymConst{\HOLTokenExists{}}\HOLBoundVar{E\sb{\mathrm{1}}}. \HOLBoundVar{a\sb{\mathrm{0}}} --\HOLBoundVar{u}-> \HOLBoundVar{E\sb{\mathrm{1}}} \HOLSymConst{\HOLTokenConj{}} \HOLFreeVar{STRONG\HOLTokenUnderscore{}EQ\sp{\prime}} \HOLBoundVar{E\sb{\mathrm{1}}} \HOLBoundVar{E\sb{\mathrm{2}}}) \HOLSymConst{\HOLTokenImp{}}
   \HOLSymConst{\HOLTokenForall{}}\HOLBoundVar{a\sb{\mathrm{0}}} \HOLBoundVar{a\sb{\mathrm{1}}}. \HOLFreeVar{STRONG\HOLTokenUnderscore{}EQ\sp{\prime}} \HOLBoundVar{a\sb{\mathrm{0}}} \HOLBoundVar{a\sb{\mathrm{1}}} \HOLSymConst{\HOLTokenImp{}} \HOLConst{STRONG_EQ} \HOLBoundVar{a\sb{\mathrm{0}}} \HOLBoundVar{a\sb{\mathrm{1}}}
STRONG_EQ_cases:
\HOLTokenTurnstile{} \HOLConst{STRONG_EQ} \HOLFreeVar{a\sb{\mathrm{0}}} \HOLFreeVar{a\sb{\mathrm{1}}} \HOLSymConst{\HOLTokenEquiv{}}
   \HOLSymConst{\HOLTokenForall{}}\HOLBoundVar{u}.
     (\HOLSymConst{\HOLTokenForall{}}\HOLBoundVar{E\sb{\mathrm{1}}}. \HOLFreeVar{a\sb{\mathrm{0}}} --\HOLBoundVar{u}-> \HOLBoundVar{E\sb{\mathrm{1}}} \HOLSymConst{\HOLTokenImp{}} \HOLSymConst{\HOLTokenExists{}}\HOLBoundVar{E\sb{\mathrm{2}}}. \HOLFreeVar{a\sb{\mathrm{1}}} --\HOLBoundVar{u}-> \HOLBoundVar{E\sb{\mathrm{2}}} \HOLSymConst{\HOLTokenConj{}} \HOLConst{STRONG_EQ} \HOLBoundVar{E\sb{\mathrm{1}}} \HOLBoundVar{E\sb{\mathrm{2}}}) \HOLSymConst{\HOLTokenConj{}}
     \HOLSymConst{\HOLTokenForall{}}\HOLBoundVar{E\sb{\mathrm{2}}}. \HOLFreeVar{a\sb{\mathrm{1}}} --\HOLBoundVar{u}-> \HOLBoundVar{E\sb{\mathrm{2}}} \HOLSymConst{\HOLTokenImp{}} \HOLSymConst{\HOLTokenExists{}}\HOLBoundVar{E\sb{\mathrm{1}}}. \HOLFreeVar{a\sb{\mathrm{0}}} --\HOLBoundVar{u}-> \HOLBoundVar{E\sb{\mathrm{1}}} \HOLSymConst{\HOLTokenConj{}} \HOLConst{STRONG_EQ} \HOLBoundVar{E\sb{\mathrm{1}}} \HOLBoundVar{E\sb{\mathrm{2}}}
\end{alltt}
The first theorem is the original rules appearing in the
definition. Roughly speaking, it's kind of rules for building a
bisimulation relation in forward way, however this is impossible
because of the lack of base rules (which exists in most inductive
relation). And it's not original in this case, since it can be derived
from the last theorem \texttt{STRONG_EQ_cases} (RHS $\Rightarrow$ LHS).

The second theorem is the co-induction principle, it says, for what
ever relation which satisfy the rules, that relation must be contained
in strong equivalence. In another word, it make sure the target
relation is the maximal relation containing all others.

The purpose of the last theorem (also called ``cases'' theorem), is to make sure the target relation
is indeed a fixed point of the function $F$ built by the given
rules. However, it
doesn't give any information about the size of such a fixed point. In
general, if the geatest fixed point and least fixed point doesn't
coincide, without the restriction by co-induction theorem, the rest
two theorems will not give a precise definition for that relation.
For strong equivalence, we already know that, the least fixed point of
$F$ is empty relation $\emptyset$, and the great fixed point is the
strong equivalance $\sim$. And in fact, the ``cases'' theorem has
``defined'' a relation which lies in the middle of the greatest and
least fixed point. To see why this argument is true, we found 
this theorem as an equation could be used as a possible definition of
strong equivalence: (c.f. p. 49 of \cite{Gorrieri:2015jt})
\begin{definition}
Define \emph{recursively} a new behavioral relation $\sim' \in Q
\times Q$ as follows: $q_1 \sim' q_2$ \emph{if and only if} for all
$\mu \in A$
\begin{itemize}
\item $\forall q_1' \text{ such that } q_1
  \overset{\mu}{\longrightarrow} q_1', \exists q_2' \text{ such that }
  q_2\overset{\mu}{\longrightarrow} q_2' \text{ and } q_1' \sim'
  q_2'$,
\item $\forall q_2' \text{ such that } q_2
  \overset{\mu}{\longrightarrow} q_2', \exists q_1' \text{ such that }
  q_1 \overset{\mu}{\longrightarrow} q_1' \text{ and } q_1' \sim'
  q_2'$.
\end{itemize}
\end{definition}
This is exactly the same as above ``cases'' theorem if the theorem
were used as a definition of strong equivalence. Robin Milner calls
this theorem the `` property (*)'' of strong
equivalence. (c.f. p.88 of \cite{Milner:2017tw}) But as Prof. Gorrieri's book \cite{Gorrieri:2015jt}
already told with examples: ``this does not identify a unique
relation, as many different relations satisfy this recursive
definition.'', and the fact that any mathematical (or logic)
definitions must precisely specify the targeting object (unless the
possible covered range itself is a targeting object).

But why the recursive definition failed to define a largest
bisimulation (i.e. strong equivalence)?
The textbooks didn't give a clear answer, but in the view of theorem
proving, now it's quite clear:
such a recursive definition can only restrict the target relation into the range of all
fixed points, while it's the co-induction thereom who finally restricts the
target relation to the greatest solution. Without any of them, the
solution will not be unique (thus not a valid mathematical definition).

Now we prove the old (\texttt{STRONG_EQUIV}, $\sim$) and new
definition (\texttt{STRONG_EQ}) of strong equivalence are
equivalent, i.e.
\begin{alltt}
STR_EQUIV_TO_STR_EQ:
\HOLTokenTurnstile{} \HOLFreeVar{E} \HOLSymConst{\HOLTokenStrongEquiv} \HOLFreeVar{E\sp{\prime}} \HOLSymConst{\HOLTokenEquiv{}} \HOLConst{STRONG_EQ} \HOLFreeVar{E} \HOLFreeVar{E\sp{\prime}}
\end{alltt}
The proof of above theorem is the result when combining the proof for
each directions:
\begin{alltt}
STR_EQ_IMP_STR_EQUIV:
\HOLTokenTurnstile{} \HOLConst{STRONG_EQ} \HOLFreeVar{E} \HOLFreeVar{E\sp{\prime}} \HOLSymConst{\HOLTokenImp{}} \HOLFreeVar{E} \HOLSymConst{\HOLTokenStrongEquiv} \HOLFreeVar{E\sp{\prime}}
STR_EQUIV_IMP_STR_EQ:
\HOLTokenTurnstile{} \HOLFreeVar{E} \HOLSymConst{\HOLTokenStrongEquiv} \HOLFreeVar{E\sp{\prime}} \HOLSymConst{\HOLTokenImp{}} \HOLConst{STRONG_EQ} \HOLFreeVar{E} \HOLFreeVar{E\sp{\prime}}
\end{alltt}
The direction from the co-inductively defined \texttt{STRONG_EQ} to
traditionally defined \texttt{STRONG_EQUIV} is relatively easy, the
proof only depends on the definition of \texttt{STRONG_EQUIV} and the
fact that \texttt{STRONG_EQ} is also a \texttt{STRONG_BISIM} relation:
\begin{alltt}
STR_EQ_IS_STR_BISIM:
\HOLTokenTurnstile{} \HOLConst{STRONG_BISIM} \HOLConst{STRONG_EQ}
\end{alltt}
which can be easily proved by comparing the definition of
\texttt{STRONG_BISIM} and the ``cases'' theorem generated from the
co-inductively defined \texttt{STRONG_EQ}. Thus the maximality of
strong equivalence is not needed.

The proof of the other direction, instead, must use the co-induction
theorem \texttt{STRONG_EQ_coind} and the ``property (*)'' of
\texttt{STRONG_EQUIV} mentioned once previouly:
\begin{alltt}
PROPERTY_STAR:
\HOLTokenTurnstile{} \HOLFreeVar{E} \HOLSymConst{\HOLTokenStrongEquiv} \HOLFreeVar{E\sp{\prime}} \HOLSymConst{\HOLTokenEquiv{}}
   \HOLSymConst{\HOLTokenForall{}}\HOLBoundVar{u}.
     (\HOLSymConst{\HOLTokenForall{}}\HOLBoundVar{E\sb{\mathrm{1}}}. \HOLFreeVar{E} --\HOLBoundVar{u}-> \HOLBoundVar{E\sb{\mathrm{1}}} \HOLSymConst{\HOLTokenImp{}} \HOLSymConst{\HOLTokenExists{}}\HOLBoundVar{E\sb{\mathrm{2}}}. \HOLFreeVar{E\sp{\prime}} --\HOLBoundVar{u}-> \HOLBoundVar{E\sb{\mathrm{2}}} \HOLSymConst{\HOLTokenConj{}} \HOLBoundVar{E\sb{\mathrm{1}}} \HOLSymConst{\HOLTokenStrongEquiv} \HOLBoundVar{E\sb{\mathrm{2}}}) \HOLSymConst{\HOLTokenConj{}}
     \HOLSymConst{\HOLTokenForall{}}\HOLBoundVar{E\sb{\mathrm{2}}}. \HOLFreeVar{E\sp{\prime}} --\HOLBoundVar{u}-> \HOLBoundVar{E\sb{\mathrm{2}}} \HOLSymConst{\HOLTokenImp{}} \HOLSymConst{\HOLTokenExists{}}\HOLBoundVar{E\sb{\mathrm{1}}}. \HOLFreeVar{E} --\HOLBoundVar{u}-> \HOLBoundVar{E\sb{\mathrm{1}}} \HOLSymConst{\HOLTokenConj{}} \HOLBoundVar{E\sb{\mathrm{1}}} \HOLSymConst{\HOLTokenStrongEquiv} \HOLBoundVar{E\sb{\mathrm{2}}}
\end{alltt}
The proof of above ``property (*)'' is similar with previous steps,
but it's not trivial. Here we omit the details (the reader can always
check the proof scripts for all details). For just one time we show how
co-induction theorem is used to prove
\texttt{STR_EQUIV_IMP_STR_EQ} and we replay this proof by HOL's
interactive proof manager\footnote{in the following quote text, the
  leading \texttt{>} is the prompt of HOL running in PolyML, the
  function \texttt{g} puts an initial goal into the proof manager, and
  the function \texttt{e} applies tacticals to current goal}:
\begin{lstlisting}
> g `!E E'. STRONG_EQUIV E E' ==> STRONG_EQ E E'`;
val it =
   Proof manager status: 1 proof.
1. Incomplete goalstack:
     Initial goal:

     !E E'. E ~ E' ==> STRONG_EQ E E'
:
   proofs
\end{lstlisting}
Now we're going to apply the co-induction theorem:
\begin{alltt}
\HOLTokenTurnstile{} (\HOLSymConst{\HOLTokenForall{}}\HOLBoundVar{a\sb{\mathrm{0}}} \HOLBoundVar{a\sb{\mathrm{1}}}.
      \HOLFreeVar{STRONG\HOLTokenUnderscore{}EQ\sp{\prime}} \HOLBoundVar{a\sb{\mathrm{0}}} \HOLBoundVar{a\sb{\mathrm{1}}} \HOLSymConst{\HOLTokenImp{}}
      \HOLSymConst{\HOLTokenForall{}}\HOLBoundVar{u}.
        (\HOLSymConst{\HOLTokenForall{}}\HOLBoundVar{E\sb{\mathrm{1}}}.
           \HOLBoundVar{a\sb{\mathrm{0}}} --\HOLBoundVar{u}-> \HOLBoundVar{E\sb{\mathrm{1}}} \HOLSymConst{\HOLTokenImp{}} \HOLSymConst{\HOLTokenExists{}}\HOLBoundVar{E\sb{\mathrm{2}}}. \HOLBoundVar{a\sb{\mathrm{1}}} --\HOLBoundVar{u}-> \HOLBoundVar{E\sb{\mathrm{2}}} \HOLSymConst{\HOLTokenConj{}} \HOLFreeVar{STRONG\HOLTokenUnderscore{}EQ\sp{\prime}} \HOLBoundVar{E\sb{\mathrm{1}}} \HOLBoundVar{E\sb{\mathrm{2}}}) \HOLSymConst{\HOLTokenConj{}}
        \HOLSymConst{\HOLTokenForall{}}\HOLBoundVar{E\sb{\mathrm{2}}}.
          \HOLBoundVar{a\sb{\mathrm{1}}} --\HOLBoundVar{u}-> \HOLBoundVar{E\sb{\mathrm{2}}} \HOLSymConst{\HOLTokenImp{}} \HOLSymConst{\HOLTokenExists{}}\HOLBoundVar{E\sb{\mathrm{1}}}. \HOLBoundVar{a\sb{\mathrm{0}}} --\HOLBoundVar{u}-> \HOLBoundVar{E\sb{\mathrm{1}}} \HOLSymConst{\HOLTokenConj{}} \HOLFreeVar{STRONG\HOLTokenUnderscore{}EQ\sp{\prime}} \HOLBoundVar{E\sb{\mathrm{1}}} \HOLBoundVar{E\sb{\mathrm{2}}}) \HOLSymConst{\HOLTokenImp{}}
   \HOLSymConst{\HOLTokenForall{}}\HOLBoundVar{a\sb{\mathrm{0}}} \HOLBoundVar{a\sb{\mathrm{1}}}. \HOLFreeVar{STRONG\HOLTokenUnderscore{}EQ\sp{\prime}} \HOLBoundVar{a\sb{\mathrm{0}}} \HOLBoundVar{a\sb{\mathrm{1}}} \HOLSymConst{\HOLTokenImp{}} \HOLConst{STRONG_EQ} \HOLBoundVar{a\sb{\mathrm{0}}} \HOLBoundVar{a\sb{\mathrm{1}}}
\end{alltt}
The tactical for applying such (co)induction theorems in HOL is to
reduces the goal using a supplied implication, with higher-order
matching, this tactical is called \texttt{HO_MATCH_MP_TAC}:
\begin{lstlisting}
> e (HO_MATCH_MP_TAC STRONG_EQ_coind);
OK..
1 subgoal:
val it =
   
!E E'.
  E ~ E' ==>
  !u.
    (!E1. E --u-> E1 ==> ?E2. E' --u-> E2 /\ E1 ~ E2) /\
    !E2. E' --u-> E2 ==> ?E1. E --u-> E1 /\ E1 ~ E2
:
   ?.proof
\end{lstlisting}
The rest steps is to use the ``property (*)'' to rewrite the right
side of the implication:
\begin{lstlisting}
> e (PURE_ONCE_REWRITE_TAC [GSYM PROPERTY_STAR]);
OK..
1 subgoal:
val it =
   
!E E'. E ~ E' ==> E ~ E'

:
   ?.proof
\end{lstlisting}
Now things get very clear, a simple rewrite with boolean theorems will
solve the goal easily:
\begin{lstlisting}
> e (RW_TAC bool_ss []);
OK..

Goal proved.
|- !E E'. E ~ E' ==> E ~ E'

Goal proved.
|- !E E'.
     E ~ E' ==>
     !u.
       (!E1. E --u-> E1 ==> ?E2. E' --u-> E2 /\ E1 ~ E2) /\
       !E2. E' --u-> E2 ==> ?E1. E --u-> E1 /\ E1 ~ E2
val it =
   Initial goal proved.
|- !E E'. E ~ E' ==> STRONG_EQ E E':
   ?.proof
\end{lstlisting}
Combining all the step together, a single Standard ML function in the
proof script can be written to finish the proof and store the theorem
with a name:
\begin{lstlisting}
val STR_EQUIV_IMP_STR_EQ = store_thm (
   "STR_EQUIV_IMP_STR_EQ",
      ``!E E'. STRONG_EQUIV E E' ==> STRONG_EQ E E'``,
    HO_MATCH_MP_TAC STRONG_EQ_coind (* co-induction principle used here! *)
 >> REPEAT GEN_TAC
 >> PURE_ONCE_REWRITE_TAC [GSYM PROPERTY_STAR]
 >> RW_TAC bool_ss []);
\end{lstlisting}

\subsection{Weak transition and weak equivalence}

The formalization of weak bisimulation, together with weak (and rooted) equivalence (also
called ``observation equivalence'' and ``observation congruence'' in
old books) is minimal in this project. In this part, the main purpose
is to define the weak equivalence co-inductively \emph{first} and then
prove the traditional definition (like \texttt{STRONG_EQUIV}) as a
theorem. We wants to convince the reader that, by using HOL's
coinduction facility, it's much easier to get the same set of theorems
like those for strong equivalence. These works are not part of the old
CCS formalization in HOL88, it belongs to the author.\footnote{Of course, we can also
  rewrite the proof scripts for strong equivalence and fully benefit
  from HOL's coinductive relation facility, but this is not very
  useful. The central idea of theorem proving is, once a theorem is
  successfully proved, its statement can be saved into disk for later
  use without the need to run the proof again everytime when it's
  used. And the proof steps are not saved at all. Thus, the same
  theorem proved by different methods, when they were saved into disk,
  there's absolutely no difference except for their names.  On the
  other side, we want to keep the old definition for at least strong
  equivalence, because it's a literature formalization of the definitions
  in standard textbooks.}

There're multiple ways to define the concept of weak transitions used
in the defintion of weak bisimulation. In early approach like Milner's
book, the first step is to define a \texttt{EPS} relation, which indicates that between two
processes there's nothing but zero or more $\tau$ transitions. In HOL,
this can be defined through a non-recursive inductive relation and the
RTC (reflexitive transitive closure) on top of it:
\begin{alltt}
\HOLTokenTurnstile{} \HOLConst{EPS1} \HOLFreeVar{a\sb{\mathrm{0}}} \HOLFreeVar{a\sb{\mathrm{1}}} \HOLSymConst{\HOLTokenEquiv{}} \HOLFreeVar{a\sb{\mathrm{0}}} --\HOLSymConst{\ensuremath{\tau}}-> \HOLFreeVar{a\sb{\mathrm{1}}}
\HOLTokenTurnstile{} \HOLConst{EPS} \HOLSymConst{=} \HOLConst{EPS1}\HOLSymConst{\HOLTokenSupStar{}}
\end{alltt}
Once we have the \HOLinline{\HOLConst{EPS}} relation, the weak transition can be defined
but a normal transition wrapped with two \HOLinline{\HOLConst{EPS}} transitions:
\begin{alltt}
\HOLTokenTurnstile{} \HOLFreeVar{a\sb{\mathrm{0}}} ==\HOLFreeVar{a\sb{\mathrm{1}}}=>> \HOLFreeVar{a\sb{\mathrm{2}}} \HOLSymConst{\HOLTokenEquiv{}}
   \HOLSymConst{\HOLTokenExists{}}\HOLBoundVar{E\sb{\mathrm{1}}} \HOLBoundVar{E\sb{\mathrm{2}}}. \HOLConst{EPS} \HOLFreeVar{a\sb{\mathrm{0}}} \HOLBoundVar{E\sb{\mathrm{1}}} \HOLSymConst{\HOLTokenConj{}} \HOLBoundVar{E\sb{\mathrm{1}}} --\HOLFreeVar{a\sb{\mathrm{1}}}-> \HOLBoundVar{E\sb{\mathrm{2}}} \HOLSymConst{\HOLTokenConj{}} \HOLConst{EPS} \HOLBoundVar{E\sb{\mathrm{2}}} \HOLFreeVar{a\sb{\mathrm{2}}}
\end{alltt}

Modern textbooks like \cite{Gorrieri:2015jt} directly uses ``weak trace''
transition for definining weak bisimulation, in which there's only one
action in the trace. Here are the definition of weak trace:
\begin{definition}{(Weak trace)}
For any LTS $TS = (Q, A \cup {\tau}, \rightarrow)$, where $\tau \notin
A$, define relation $\Longrightarrow \subset Q \times A^* \times Q$ as
the \emph{weak} reflexive and transitive closure of $\rightarrow$,
i.e., as the least relation induced by the following axiom and rules,
where $\epsilon$ is the empty trace:
\begin{multicols}{4}
\begin{prooftree}
\AxiomC{$q_1 \overset{\alpha}{\longrightarrow} q_2$}
\UnaryInfC{$q_1 \overset{\alpha}{\Longrightarrow} q_2$}
\end{prooftree}

\begin{prooftree}
\AxiomC{$q_1 \overset{\tau}{\longrightarrow} q_2$}
\UnaryInfC{$q_1 \overset{\epsilon}{\Longrightarrow} q_2$}
\end{prooftree}

\begin{prooftree}
\AxiomC{}
\UnaryInfC{$q \overset{\epsilon}{\Longrightarrow} q$}
\end{prooftree}

\begin{prooftree}
\AxiomC{$q_1 \overset{\sigma_1}{\Longrightarrow} q_2$}
\AxiomC{$q_2 \overset{\sigma_2}{\Longrightarrow} q_3$}
\BinaryInfC{$q_1 \overset{\sigma_1 \sigma_2}{\Longrightarrow} q_3$}
\end{prooftree}
\end{multicols}
\end{definition}

In HOL, we can use a list of \HOLinline{\HOLTyOp{Label}} to represent the trace (thus
there's naturally no $\tau$ in the list, as $\tau$ is not part of the
type \HOLinline{\HOLTyOp{Label}} but \HOLinline{\HOLTyOp{Action}}) and empty list can be seen as
the $\epsilon$. As the result, the relation \HOLinline{\HOLConst{WEAK_TRACE}} has
type \HOLinline{\HOLTyOp{CCS} -> \HOLTyOp{Label} \HOLTyOp{list} -> \HOLTyOp{CCS} -> \HOLTyOp{bool}}. Below is the ``rules'' theorem generated by
\texttt{Hol_reln} command:
\begin{alltt}
\HOLTokenTurnstile{} (\HOLSymConst{\HOLTokenForall{}}\HOLBoundVar{E}. \HOLConst{WEAK_TRACE} \HOLBoundVar{E} \HOLSymConst{\HOLepsilon} \HOLBoundVar{E}) \HOLSymConst{\HOLTokenConj{}}
   (\HOLSymConst{\HOLTokenForall{}}\HOLBoundVar{E} \HOLBoundVar{E\sp{\prime}}. \HOLBoundVar{E} --\HOLSymConst{\ensuremath{\tau}}-> \HOLBoundVar{E\sp{\prime}} \HOLSymConst{\HOLTokenImp{}} \HOLConst{WEAK_TRACE} \HOLBoundVar{E} \HOLSymConst{\HOLepsilon} \HOLBoundVar{E\sp{\prime}}) \HOLSymConst{\HOLTokenConj{}}
   (\HOLSymConst{\HOLTokenForall{}}\HOLBoundVar{E} \HOLBoundVar{E\sp{\prime}} \HOLBoundVar{l}. \HOLBoundVar{E} --\HOLConst{label} \HOLBoundVar{l}-> \HOLBoundVar{E\sp{\prime}} \HOLSymConst{\HOLTokenImp{}} \HOLConst{WEAK_TRACE} \HOLBoundVar{E} [\HOLBoundVar{l}] \HOLBoundVar{E\sp{\prime}}) \HOLSymConst{\HOLTokenConj{}}
   \HOLSymConst{\HOLTokenForall{}}\HOLBoundVar{E\sb{\mathrm{1}}} \HOLBoundVar{E\sb{\mathrm{2}}} \HOLBoundVar{E\sb{\mathrm{3}}} \HOLBoundVar{l\sb{\mathrm{1}}} \HOLBoundVar{l\sb{\mathrm{2}}}.
     \HOLConst{WEAK_TRACE} \HOLBoundVar{E\sb{\mathrm{1}}} \HOLBoundVar{l\sb{\mathrm{1}}} \HOLBoundVar{E\sb{\mathrm{2}}} \HOLSymConst{\HOLTokenConj{}} \HOLConst{WEAK_TRACE} \HOLBoundVar{E\sb{\mathrm{2}}} \HOLBoundVar{l\sb{\mathrm{2}}} \HOLBoundVar{E\sb{\mathrm{3}}} \HOLSymConst{\HOLTokenImp{}}
     \HOLConst{WEAK_TRACE} \HOLBoundVar{E\sb{\mathrm{1}}} (\HOLBoundVar{l\sb{\mathrm{1}}} \HOLSymConst{++} \HOLBoundVar{l\sb{\mathrm{2}}}) \HOLBoundVar{E\sb{\mathrm{3}}}
\end{alltt}

Now we take a look at the definition of weak bisimulation:
\begin{definition}{(Weak bisimulation and weak equivalence)}
For any LTS $(Q, A \cup {\tau}, \rightarrow)$, where $\tau \notin
A$, a \emph{weak bisimulation} is a relation $R \subset Q\times Q$
such that both $R$ and its inverse $R^{-1}$ are weak simulations. More
explicitly, a weak bisimulation is a relation $R$ such that if
$(q_1,q_2) \in R$ then for all $\alpha \in A$
\begin{itemize}
\item $\forall q_1' \text{ such that }
  q_1\overset{\alpha}{\longrightarrow} q_1', \exists q_2' \text{ such
    that } q_2\overset{\alpha}{\Longrightarrow} q_2' \text{ and }
  (q_1', q_2') \in R$,
\item $\forall q_1' \text{ such that }
  q_1\overset{\tau}{\longrightarrow} q_1', \exists q_2' \text{ such
    that } q_2\overset{\epsilon}{\Longrightarrow} q_2' \text{ and }
  (q_1', q_2') \in R$,
\end{itemize}
and, summetrically,
\begin{itemize}
\item $\forall q_2' \text{ such that }
  q_2\overset{\alpha}{\longrightarrow} q_2', \exists q_1' \text{ such
    that } q_1\overset{\alpha}{\Longrightarrow} q_1' \text{ and }
  (q_1', q_2') \in R$,
\item $\forall q_2' \text{ such that }
  q_2\overset{\tau}{\longrightarrow} q_2', \exists q_1' \text{ such
    that } q_1\overset{\epsilon}{\Longrightarrow} q_1' \text{ and }
  (q_1', q_2') \in R$.
\end{itemize}
States $q$ and $q'$ are \emph{weakly bisimilar} (or \emph{weak
  bisimulation equivalent}), denoted with $q \approx q'$, if there
exists a weak bisimulation $R$ such that $(q,q') \in R$.
\end{definition}

There's no big problem to use all weak traces in above definition, as
long as we limit the number of labels in the trace to just one. The
real difficulty happens when we try to further define the \emph{rooted
  weak bisimilarity} on top of weak equivalence, in which an auxiliary
relation $q \overset{\tau}{\Longrightarrow} q'$ must be defined as
\begin{equation}
q \overset{\tau}{\Longrightarrow} q' \text{ if and only if }
\exists q_1, q_2,\enspace q \overset{\epsilon}{\Longrightarrow} q_1
\overset{\tau}{\longrightarrow} q_2
\overset{\epsilon}{\Longrightarrow} q'.
\end{equation}
And the definition of rooted weak bisimilarity (in Milner's book it's
also called ``observation congruence''), noticed that it's not recursive:
\begin{definition}{(Rooted weak bisimilarity)}
Given an LTS $(Q, A \cup {\tau}, \rightarrow)$, two states $q_1$ and
$q_2$ are rooted weak bisimilar, denoted $q_1 \approx^c q_2$, if for
all $\mu \in A \cup \{\tau\}$
\begin{itemize}
\item $\forall q_1' \text{ such that }
  q_1\overset{\mu}{\longrightarrow} q_1', \exists q_2' \text{ such
    that } q_2\overset{\mu}{\Longrightarrow} q_2' \text{ and }
  q_1' \approx q_2'$,
\item $\forall q_2' \text{ such that }
  q_2\overset{\mu}{\longrightarrow} q_2', \exists q_1' \text{ such
    that } q_1\overset{\mu}{\Longrightarrow} q_1' \text{ and }
  q_1' \approx q_2'$.
\end{itemize}
\end{definition}

But there's a type error here: $\tau$ is not a legal trace, thus a
term like \texttt{WEAK_TRACE E [tau]
  E'} is illegal with the existing definition of
\texttt{WEAK_TRACE}. If we enlarge the type of traces to
\HOLinline{\HOLTyOp{Action} \HOLTyOp{list}}, then invalid traces like $\sigma_1 \tau
\sigma_2$ will become legal in HOL terms.  One dirty solution is to
define another relation \texttt{WEAK_TRACE'} which takes single \HOLinline{\HOLTyOp{Action}}
having the above auxiliary
relation as part of is definition. But this actually coincides with
\HOLinline{\HOLConst{WEAK_TRANS}}.

Since the use of \texttt{WEAK_TRACE} will cause theorems from HOL's
\texttt{listTheory} being used to handle the list of labels. To
simplify things, in all these weak bisimulation
variants, we only use \HOLinline{\HOLConst{WEAK_TRANS}} and \HOLinline{\HOLConst{EPS}}. Here is the
definition of \HOLinline{\HOLConst{WEAK_BISIM}} (not recursive) in HOL:
\begin{alltt}
\HOLTokenTurnstile{} \HOLConst{WEAK_BISIM} \HOLFreeVar{Wbsm} \HOLSymConst{\HOLTokenEquiv{}}
   \HOLSymConst{\HOLTokenForall{}}\HOLBoundVar{E} \HOLBoundVar{E\sp{\prime}}.
     \HOLFreeVar{Wbsm} \HOLBoundVar{E} \HOLBoundVar{E\sp{\prime}} \HOLSymConst{\HOLTokenImp{}}
     (\HOLSymConst{\HOLTokenForall{}}\HOLBoundVar{l}.
        (\HOLSymConst{\HOLTokenForall{}}\HOLBoundVar{E\sb{\mathrm{1}}}.
           \HOLBoundVar{E} --\HOLConst{label} \HOLBoundVar{l}-> \HOLBoundVar{E\sb{\mathrm{1}}} \HOLSymConst{\HOLTokenImp{}}
           \HOLSymConst{\HOLTokenExists{}}\HOLBoundVar{E\sb{\mathrm{2}}}. \HOLBoundVar{E\sp{\prime}} ==\HOLConst{label} \HOLBoundVar{l}=>> \HOLBoundVar{E\sb{\mathrm{2}}} \HOLSymConst{\HOLTokenConj{}} \HOLFreeVar{Wbsm} \HOLBoundVar{E\sb{\mathrm{1}}} \HOLBoundVar{E\sb{\mathrm{2}}}) \HOLSymConst{\HOLTokenConj{}}
        \HOLSymConst{\HOLTokenForall{}}\HOLBoundVar{E\sb{\mathrm{2}}}.
          \HOLBoundVar{E\sp{\prime}} --\HOLConst{label} \HOLBoundVar{l}-> \HOLBoundVar{E\sb{\mathrm{2}}} \HOLSymConst{\HOLTokenImp{}}
          \HOLSymConst{\HOLTokenExists{}}\HOLBoundVar{E\sb{\mathrm{1}}}. \HOLBoundVar{E} ==\HOLConst{label} \HOLBoundVar{l}=>> \HOLBoundVar{E\sb{\mathrm{1}}} \HOLSymConst{\HOLTokenConj{}} \HOLFreeVar{Wbsm} \HOLBoundVar{E\sb{\mathrm{1}}} \HOLBoundVar{E\sb{\mathrm{2}}}) \HOLSymConst{\HOLTokenConj{}}
     (\HOLSymConst{\HOLTokenForall{}}\HOLBoundVar{E\sb{\mathrm{1}}}. \HOLBoundVar{E} --\HOLSymConst{\ensuremath{\tau}}-> \HOLBoundVar{E\sb{\mathrm{1}}} \HOLSymConst{\HOLTokenImp{}} \HOLSymConst{\HOLTokenExists{}}\HOLBoundVar{E\sb{\mathrm{2}}}. \HOLConst{EPS} \HOLBoundVar{E\sp{\prime}} \HOLBoundVar{E\sb{\mathrm{2}}} \HOLSymConst{\HOLTokenConj{}} \HOLFreeVar{Wbsm} \HOLBoundVar{E\sb{\mathrm{1}}} \HOLBoundVar{E\sb{\mathrm{2}}}) \HOLSymConst{\HOLTokenConj{}}
     \HOLSymConst{\HOLTokenForall{}}\HOLBoundVar{E\sb{\mathrm{2}}}. \HOLBoundVar{E\sp{\prime}} --\HOLSymConst{\ensuremath{\tau}}-> \HOLBoundVar{E\sb{\mathrm{2}}} \HOLSymConst{\HOLTokenImp{}} \HOLSymConst{\HOLTokenExists{}}\HOLBoundVar{E\sb{\mathrm{1}}}. \HOLConst{EPS} \HOLBoundVar{E} \HOLBoundVar{E\sb{\mathrm{1}}} \HOLSymConst{\HOLTokenConj{}} \HOLFreeVar{Wbsm} \HOLBoundVar{E\sb{\mathrm{1}}} \HOLBoundVar{E\sb{\mathrm{2}}}
\end{alltt}
And the relation \HOLinline{\HOLConst{WEAK_EQUIV}} is co-inductively defined by HOL's
\texttt{Hol_coreln} command:
\begin{lstlisting}
val (WEAK_EQUIV_rules, WEAK_EQUIV_coind, WEAK_EQUIV_cases) = Hol_coreln `
    (!E E'.
       (!l.
         (!E1. TRANS E  (label l) E1 ==>
               (?E2. WEAK_TRANS E' (label l) E2 /\ WEAK_EQUIV E1 E2)) /\
         (!E2. TRANS E' (label l) E2 ==>
               (?E1. WEAK_TRANS E  (label l) E1 /\ WEAK_EQUIV E1 E2))) /\
       (!E1. TRANS E  tau E1 ==> (?E2. EPS E' E2 /\ WEAK_EQUIV E1 E2)) /\
       (!E2. TRANS E' tau E2 ==> (?E1. EPS E  E1 /\ WEAK_EQUIV E1 E2))
      ==> WEAK_EQUIV E E')`;
\end{lstlisting}
Like for the strong equivalence, the above command generates three theorems
which fully characteristics the weak equivalence relation:
\begin{enumerate}
\item The (forward) rules for weak equivalance:
\begin{alltt}
\HOLTokenTurnstile{} (\HOLSymConst{\HOLTokenForall{}}\HOLBoundVar{l}.
      (\HOLSymConst{\HOLTokenForall{}}\HOLBoundVar{E\sb{\mathrm{1}}}.
         \HOLFreeVar{E} --\HOLConst{label} \HOLBoundVar{l}-> \HOLBoundVar{E\sb{\mathrm{1}}} \HOLSymConst{\HOLTokenImp{}}
         \HOLSymConst{\HOLTokenExists{}}\HOLBoundVar{E\sb{\mathrm{2}}}. \HOLFreeVar{E\sp{\prime}} ==\HOLConst{label} \HOLBoundVar{l}=>> \HOLBoundVar{E\sb{\mathrm{2}}} \HOLSymConst{\HOLTokenConj{}} \HOLBoundVar{E\sb{\mathrm{1}}} \HOLSymConst{\HOLTokenWeakEquiv} \HOLBoundVar{E\sb{\mathrm{2}}}) \HOLSymConst{\HOLTokenConj{}}
      \HOLSymConst{\HOLTokenForall{}}\HOLBoundVar{E\sb{\mathrm{2}}}.
        \HOLFreeVar{E\sp{\prime}} --\HOLConst{label} \HOLBoundVar{l}-> \HOLBoundVar{E\sb{\mathrm{2}}} \HOLSymConst{\HOLTokenImp{}} \HOLSymConst{\HOLTokenExists{}}\HOLBoundVar{E\sb{\mathrm{1}}}. \HOLFreeVar{E} ==\HOLConst{label} \HOLBoundVar{l}=>> \HOLBoundVar{E\sb{\mathrm{1}}} \HOLSymConst{\HOLTokenConj{}} \HOLBoundVar{E\sb{\mathrm{1}}} \HOLSymConst{\HOLTokenWeakEquiv} \HOLBoundVar{E\sb{\mathrm{2}}}) \HOLSymConst{\HOLTokenConj{}}
   (\HOLSymConst{\HOLTokenForall{}}\HOLBoundVar{E\sb{\mathrm{1}}}. \HOLFreeVar{E} --\HOLSymConst{\ensuremath{\tau}}-> \HOLBoundVar{E\sb{\mathrm{1}}} \HOLSymConst{\HOLTokenImp{}} \HOLSymConst{\HOLTokenExists{}}\HOLBoundVar{E\sb{\mathrm{2}}}. \HOLConst{EPS} \HOLFreeVar{E\sp{\prime}} \HOLBoundVar{E\sb{\mathrm{2}}} \HOLSymConst{\HOLTokenConj{}} \HOLBoundVar{E\sb{\mathrm{1}}} \HOLSymConst{\HOLTokenWeakEquiv} \HOLBoundVar{E\sb{\mathrm{2}}}) \HOLSymConst{\HOLTokenConj{}}
   (\HOLSymConst{\HOLTokenForall{}}\HOLBoundVar{E\sb{\mathrm{2}}}. \HOLFreeVar{E\sp{\prime}} --\HOLSymConst{\ensuremath{\tau}}-> \HOLBoundVar{E\sb{\mathrm{2}}} \HOLSymConst{\HOLTokenImp{}} \HOLSymConst{\HOLTokenExists{}}\HOLBoundVar{E\sb{\mathrm{1}}}. \HOLConst{EPS} \HOLFreeVar{E} \HOLBoundVar{E\sb{\mathrm{1}}} \HOLSymConst{\HOLTokenConj{}} \HOLBoundVar{E\sb{\mathrm{1}}} \HOLSymConst{\HOLTokenWeakEquiv} \HOLBoundVar{E\sb{\mathrm{2}}}) \HOLSymConst{\HOLTokenImp{}}
   \HOLFreeVar{E} \HOLSymConst{\HOLTokenWeakEquiv} \HOLFreeVar{E\sp{\prime}}
\end{alltt}
\item The co-induction theorem which assert the maximality of the relation:
\begin{alltt}
\HOLTokenTurnstile{} (\HOLSymConst{\HOLTokenForall{}}\HOLBoundVar{a\sb{\mathrm{0}}} \HOLBoundVar{a\sb{\mathrm{1}}}.
      \HOLFreeVar{WEAK\HOLTokenUnderscore{}EQUIV\sp{\prime}} \HOLBoundVar{a\sb{\mathrm{0}}} \HOLBoundVar{a\sb{\mathrm{1}}} \HOLSymConst{\HOLTokenImp{}}
      (\HOLSymConst{\HOLTokenForall{}}\HOLBoundVar{l}.
         (\HOLSymConst{\HOLTokenForall{}}\HOLBoundVar{E\sb{\mathrm{1}}}.
            \HOLBoundVar{a\sb{\mathrm{0}}} --\HOLConst{label} \HOLBoundVar{l}-> \HOLBoundVar{E\sb{\mathrm{1}}} \HOLSymConst{\HOLTokenImp{}}
            \HOLSymConst{\HOLTokenExists{}}\HOLBoundVar{E\sb{\mathrm{2}}}. \HOLBoundVar{a\sb{\mathrm{1}}} ==\HOLConst{label} \HOLBoundVar{l}=>> \HOLBoundVar{E\sb{\mathrm{2}}} \HOLSymConst{\HOLTokenConj{}} \HOLFreeVar{WEAK\HOLTokenUnderscore{}EQUIV\sp{\prime}} \HOLBoundVar{E\sb{\mathrm{1}}} \HOLBoundVar{E\sb{\mathrm{2}}}) \HOLSymConst{\HOLTokenConj{}}
         \HOLSymConst{\HOLTokenForall{}}\HOLBoundVar{E\sb{\mathrm{2}}}.
           \HOLBoundVar{a\sb{\mathrm{1}}} --\HOLConst{label} \HOLBoundVar{l}-> \HOLBoundVar{E\sb{\mathrm{2}}} \HOLSymConst{\HOLTokenImp{}}
           \HOLSymConst{\HOLTokenExists{}}\HOLBoundVar{E\sb{\mathrm{1}}}. \HOLBoundVar{a\sb{\mathrm{0}}} ==\HOLConst{label} \HOLBoundVar{l}=>> \HOLBoundVar{E\sb{\mathrm{1}}} \HOLSymConst{\HOLTokenConj{}} \HOLFreeVar{WEAK\HOLTokenUnderscore{}EQUIV\sp{\prime}} \HOLBoundVar{E\sb{\mathrm{1}}} \HOLBoundVar{E\sb{\mathrm{2}}}) \HOLSymConst{\HOLTokenConj{}}
      (\HOLSymConst{\HOLTokenForall{}}\HOLBoundVar{E\sb{\mathrm{1}}}.
         \HOLBoundVar{a\sb{\mathrm{0}}} --\HOLSymConst{\ensuremath{\tau}}-> \HOLBoundVar{E\sb{\mathrm{1}}} \HOLSymConst{\HOLTokenImp{}} \HOLSymConst{\HOLTokenExists{}}\HOLBoundVar{E\sb{\mathrm{2}}}. \HOLConst{EPS} \HOLBoundVar{a\sb{\mathrm{1}}} \HOLBoundVar{E\sb{\mathrm{2}}} \HOLSymConst{\HOLTokenConj{}} \HOLFreeVar{WEAK\HOLTokenUnderscore{}EQUIV\sp{\prime}} \HOLBoundVar{E\sb{\mathrm{1}}} \HOLBoundVar{E\sb{\mathrm{2}}}) \HOLSymConst{\HOLTokenConj{}}
      \HOLSymConst{\HOLTokenForall{}}\HOLBoundVar{E\sb{\mathrm{2}}}. \HOLBoundVar{a\sb{\mathrm{1}}} --\HOLSymConst{\ensuremath{\tau}}-> \HOLBoundVar{E\sb{\mathrm{2}}} \HOLSymConst{\HOLTokenImp{}} \HOLSymConst{\HOLTokenExists{}}\HOLBoundVar{E\sb{\mathrm{1}}}. \HOLConst{EPS} \HOLBoundVar{a\sb{\mathrm{0}}} \HOLBoundVar{E\sb{\mathrm{1}}} \HOLSymConst{\HOLTokenConj{}} \HOLFreeVar{WEAK\HOLTokenUnderscore{}EQUIV\sp{\prime}} \HOLBoundVar{E\sb{\mathrm{1}}} \HOLBoundVar{E\sb{\mathrm{2}}}) \HOLSymConst{\HOLTokenImp{}}
   \HOLSymConst{\HOLTokenForall{}}\HOLBoundVar{a\sb{\mathrm{0}}} \HOLBoundVar{a\sb{\mathrm{1}}}. \HOLFreeVar{WEAK\HOLTokenUnderscore{}EQUIV\sp{\prime}} \HOLBoundVar{a\sb{\mathrm{0}}} \HOLBoundVar{a\sb{\mathrm{1}}} \HOLSymConst{\HOLTokenImp{}} \HOLBoundVar{a\sb{\mathrm{0}}} \HOLSymConst{\HOLTokenWeakEquiv} \HOLBoundVar{a\sb{\mathrm{1}}}
\end{alltt}
\item The ``cases'' theorem (or ``property (*)'') for weak
  equivalence:
\begin{alltt}
\HOLTokenTurnstile{} \HOLFreeVar{a\sb{\mathrm{0}}} \HOLSymConst{\HOLTokenWeakEquiv} \HOLFreeVar{a\sb{\mathrm{1}}} \HOLSymConst{\HOLTokenEquiv{}}
   (\HOLSymConst{\HOLTokenForall{}}\HOLBoundVar{l}.
      (\HOLSymConst{\HOLTokenForall{}}\HOLBoundVar{E\sb{\mathrm{1}}}.
         \HOLFreeVar{a\sb{\mathrm{0}}} --\HOLConst{label} \HOLBoundVar{l}-> \HOLBoundVar{E\sb{\mathrm{1}}} \HOLSymConst{\HOLTokenImp{}}
         \HOLSymConst{\HOLTokenExists{}}\HOLBoundVar{E\sb{\mathrm{2}}}. \HOLFreeVar{a\sb{\mathrm{1}}} ==\HOLConst{label} \HOLBoundVar{l}=>> \HOLBoundVar{E\sb{\mathrm{2}}} \HOLSymConst{\HOLTokenConj{}} \HOLBoundVar{E\sb{\mathrm{1}}} \HOLSymConst{\HOLTokenWeakEquiv} \HOLBoundVar{E\sb{\mathrm{2}}}) \HOLSymConst{\HOLTokenConj{}}
      \HOLSymConst{\HOLTokenForall{}}\HOLBoundVar{E\sb{\mathrm{2}}}.
        \HOLFreeVar{a\sb{\mathrm{1}}} --\HOLConst{label} \HOLBoundVar{l}-> \HOLBoundVar{E\sb{\mathrm{2}}} \HOLSymConst{\HOLTokenImp{}}
        \HOLSymConst{\HOLTokenExists{}}\HOLBoundVar{E\sb{\mathrm{1}}}. \HOLFreeVar{a\sb{\mathrm{0}}} ==\HOLConst{label} \HOLBoundVar{l}=>> \HOLBoundVar{E\sb{\mathrm{1}}} \HOLSymConst{\HOLTokenConj{}} \HOLBoundVar{E\sb{\mathrm{1}}} \HOLSymConst{\HOLTokenWeakEquiv} \HOLBoundVar{E\sb{\mathrm{2}}}) \HOLSymConst{\HOLTokenConj{}}
   (\HOLSymConst{\HOLTokenForall{}}\HOLBoundVar{E\sb{\mathrm{1}}}. \HOLFreeVar{a\sb{\mathrm{0}}} --\HOLSymConst{\ensuremath{\tau}}-> \HOLBoundVar{E\sb{\mathrm{1}}} \HOLSymConst{\HOLTokenImp{}} \HOLSymConst{\HOLTokenExists{}}\HOLBoundVar{E\sb{\mathrm{2}}}. \HOLConst{EPS} \HOLFreeVar{a\sb{\mathrm{1}}} \HOLBoundVar{E\sb{\mathrm{2}}} \HOLSymConst{\HOLTokenConj{}} \HOLBoundVar{E\sb{\mathrm{1}}} \HOLSymConst{\HOLTokenWeakEquiv} \HOLBoundVar{E\sb{\mathrm{2}}}) \HOLSymConst{\HOLTokenConj{}}
   \HOLSymConst{\HOLTokenForall{}}\HOLBoundVar{E\sb{\mathrm{2}}}. \HOLFreeVar{a\sb{\mathrm{1}}} --\HOLSymConst{\ensuremath{\tau}}-> \HOLBoundVar{E\sb{\mathrm{2}}} \HOLSymConst{\HOLTokenImp{}} \HOLSymConst{\HOLTokenExists{}}\HOLBoundVar{E\sb{\mathrm{1}}}. \HOLConst{EPS} \HOLFreeVar{a\sb{\mathrm{0}}} \HOLBoundVar{E\sb{\mathrm{1}}} \HOLSymConst{\HOLTokenConj{}} \HOLBoundVar{E\sb{\mathrm{1}}} \HOLSymConst{\HOLTokenWeakEquiv} \HOLBoundVar{E\sb{\mathrm{2}}}
\end{alltt}
\end{enumerate}

Unlike in the definition of strong equivalence, our definition of
\HOLinline{\HOLConst{WEAK_EQUIV}} is unrelated to the definition of
\HOLinline{\HOLConst{WEAK_BISIM}}. But we want to show that, the textbook definition
for weak equivalence which is similar with the definition of
strong equivalence
\begin{alltt}
\HOLTokenTurnstile{} \HOLFreeVar{E} \HOLSymConst{\HOLTokenStrongEquiv} \HOLFreeVar{E\sp{\prime}} \HOLSymConst{\HOLTokenEquiv{}} \HOLSymConst{\HOLTokenExists{}}\HOLBoundVar{Bsm}. \HOLBoundVar{Bsm} \HOLFreeVar{E} \HOLFreeVar{E\sp{\prime}} \HOLSymConst{\HOLTokenConj{}} \HOLConst{STRONG_BISIM} \HOLBoundVar{Bsm}
\end{alltt}
can now be proved as a theorem:
\begin{alltt}
\HOLTokenTurnstile{} \HOLFreeVar{E} \HOLSymConst{\HOLTokenWeakEquiv} \HOLFreeVar{E\sp{\prime}} \HOLSymConst{\HOLTokenEquiv{}} \HOLSymConst{\HOLTokenExists{}}\HOLBoundVar{Wbsm}. \HOLBoundVar{Wbsm} \HOLFreeVar{E} \HOLFreeVar{E\sp{\prime}} \HOLSymConst{\HOLTokenConj{}} \HOLConst{WEAK_BISIM} \HOLBoundVar{Wbsm}
\end{alltt}
The proof is very simple, because now we have the ``property (*)'' for
free. The first step is to prove that the weak equivalence is also a
weak bisimilation relation:
\begin{alltt}
\HOLTokenTurnstile{} \HOLConst{WEAK_BISIM} \HOLConst{WEAK_EQUIV}
\end{alltt}
Then in the proof of \texttt{WEAK_EQUIV}, one direction can be easily
proved by above theorem, and other direction can also be proved easily
by co-induction theorem and first-order proof searching (using HOL's
METIS_TAC \cite{Hurd:2003wra}):
\begin{lstlisting}
(* Alternative definition of WEAK_EQUIV, similar with STRONG_EQUIV (definition).
   "Weak bisimilarity contains all weak bisimulations (thus maximal)"
 *)
val WEAK_EQUIV = store_thm ("WEAK_EQUIV",
  ``!E E'. WEAK_EQUIV E E' = (?Wbsm. Wbsm E E' /\ WEAK_BISIM Wbsm)``,
    REPEAT GEN_TAC
 >> EQ_TAC (* 2 sub-goals here *)
 >| [ (* goal 1 (of 2) *)
      DISCH_TAC \\
      EXISTS_TAC ``WEAK_EQUIV`` \\
      ASM_REWRITE_TAC [WEAK_EQUIV_IS_WEAK_BISIM],
      (* goal 2 (of 2) *)
      Q.SPEC_TAC (`E'`, `E'`) \\
      Q.SPEC_TAC (`E`, `E`) \\
      HO_MATCH_MP_TAC WEAK_EQUIV_coind \\ (* co-induction used here! *)
      METIS_TAC [WEAK_BISIM] ]);
\end{lstlisting}

As mentioned in the literature \cite{Sangiorgi:2011ut}, bisimilation
equivalence is
one of the most well-studied co-inductive relation. But since HOL (and
other theorem provers like Coq and Isabelle) started to support the
co-inductive relation features (in very recent years), the correctness of these features were
never confirmed on the (strong and weak) bisimilation equivalence
defined on CCS-like graph structures. Now in this project, we have
finally done this experiment.

Finally the rooted weak equivalence is defined in HOL as follows:
(again, not recursive)
\begin{alltt}
\HOLTokenTurnstile{} \HOLFreeVar{E} \HOLSymConst{\HOLTokenRootedWeakEquiv} \HOLFreeVar{E\sp{\prime}} \HOLSymConst{\HOLTokenEquiv{}}
   \HOLSymConst{\HOLTokenForall{}}\HOLBoundVar{u}.
     (\HOLSymConst{\HOLTokenForall{}}\HOLBoundVar{E\sb{\mathrm{1}}}. \HOLFreeVar{E} --\HOLBoundVar{u}-> \HOLBoundVar{E\sb{\mathrm{1}}} \HOLSymConst{\HOLTokenImp{}} \HOLSymConst{\HOLTokenExists{}}\HOLBoundVar{E\sb{\mathrm{2}}}. \HOLFreeVar{E\sp{\prime}} ==\HOLBoundVar{u}=>> \HOLBoundVar{E\sb{\mathrm{2}}} \HOLSymConst{\HOLTokenConj{}} \HOLBoundVar{E\sb{\mathrm{1}}} \HOLSymConst{\HOLTokenWeakEquiv} \HOLBoundVar{E\sb{\mathrm{2}}}) \HOLSymConst{\HOLTokenConj{}}
     \HOLSymConst{\HOLTokenForall{}}\HOLBoundVar{E\sb{\mathrm{2}}}. \HOLFreeVar{E\sp{\prime}} --\HOLBoundVar{u}-> \HOLBoundVar{E\sb{\mathrm{2}}} \HOLSymConst{\HOLTokenImp{}} \HOLSymConst{\HOLTokenExists{}}\HOLBoundVar{E\sb{\mathrm{1}}}. \HOLFreeVar{E} ==\HOLBoundVar{u}=>> \HOLBoundVar{E\sb{\mathrm{1}}} \HOLSymConst{\HOLTokenConj{}} \HOLBoundVar{E\sb{\mathrm{1}}} \HOLSymConst{\HOLTokenWeakEquiv} \HOLBoundVar{E\sb{\mathrm{2}}}
\end{alltt}
However, there's no theorems proven for rooted weak equivalences in
this project.

\subsection{Laws for strong equivalence}

Based on the definition of \HOLinline{\HOLConst{STRONG_EQUIV}} and SOS inference
rules for the \texttt{TRANS} relation, we have proved a large set of
theorems concerning the strong equivalence of CCS processes. Below is
a list of fundamental congruence theorems for strong equivalence:
\begin{alltt}
STRONG_EQUIV_SUBST_PREFIX:
\HOLTokenTurnstile{} \HOLFreeVar{E} \HOLSymConst{\HOLTokenStrongEquiv} \HOLFreeVar{E\sp{\prime}} \HOLSymConst{\HOLTokenImp{}} \HOLSymConst{\HOLTokenForall{}}\HOLBoundVar{u}. \HOLBoundVar{u}\HOLSymConst{..}\HOLFreeVar{E} \HOLSymConst{\HOLTokenStrongEquiv} \HOLBoundVar{u}\HOLSymConst{..}\HOLFreeVar{E\sp{\prime}}
STRONG_EQUIV_PRESD_BY_SUM:
\HOLTokenTurnstile{} \HOLFreeVar{E\sb{\mathrm{1}}} \HOLSymConst{\HOLTokenStrongEquiv} \HOLFreeVar{E\sb{\mathrm{1}}\sp{\prime}} \HOLSymConst{\HOLTokenConj{}} \HOLFreeVar{E\sb{\mathrm{2}}} \HOLSymConst{\HOLTokenStrongEquiv} \HOLFreeVar{E\sb{\mathrm{2}}\sp{\prime}} \HOLSymConst{\HOLTokenImp{}} \HOLFreeVar{E\sb{\mathrm{1}}} \HOLSymConst{+} \HOLFreeVar{E\sb{\mathrm{2}}} \HOLSymConst{\HOLTokenStrongEquiv} \HOLFreeVar{E\sb{\mathrm{1}}\sp{\prime}} \HOLSymConst{+} \HOLFreeVar{E\sb{\mathrm{2}}\sp{\prime}}
STRONG_EQUIV_PRESD_BY_PAR:
\HOLTokenTurnstile{} \HOLFreeVar{E\sb{\mathrm{1}}} \HOLSymConst{\HOLTokenStrongEquiv} \HOLFreeVar{E\sb{\mathrm{1}}\sp{\prime}} \HOLSymConst{\HOLTokenConj{}} \HOLFreeVar{E\sb{\mathrm{2}}} \HOLSymConst{\HOLTokenStrongEquiv} \HOLFreeVar{E\sb{\mathrm{2}}\sp{\prime}} \HOLSymConst{\HOLTokenImp{}} \HOLFreeVar{E\sb{\mathrm{1}}} \HOLSymConst{||} \HOLFreeVar{E\sb{\mathrm{2}}} \HOLSymConst{\HOLTokenStrongEquiv} \HOLFreeVar{E\sb{\mathrm{1}}\sp{\prime}} \HOLSymConst{||} \HOLFreeVar{E\sb{\mathrm{2}}\sp{\prime}}
STRONG_EQUIV_SUBST_RESTR:
\HOLTokenTurnstile{} \HOLFreeVar{E} \HOLSymConst{\HOLTokenStrongEquiv} \HOLFreeVar{E\sp{\prime}} \HOLSymConst{\HOLTokenImp{}} \HOLSymConst{\HOLTokenForall{}}\HOLBoundVar{L}. \HOLSymConst{\ensuremath{\nu}} \HOLBoundVar{L} \HOLFreeVar{E} \HOLSymConst{\HOLTokenStrongEquiv} \HOLSymConst{\ensuremath{\nu}} \HOLBoundVar{L} \HOLFreeVar{E\sp{\prime}}
STRONG_EQUIV_SUBST_RELAB:
\HOLTokenTurnstile{} \HOLFreeVar{E} \HOLSymConst{\HOLTokenStrongEquiv} \HOLFreeVar{E\sp{\prime}} \HOLSymConst{\HOLTokenImp{}} \HOLSymConst{\HOLTokenForall{}}\HOLBoundVar{rf}. \HOLConst{relab} \HOLFreeVar{E} \HOLBoundVar{rf} \HOLSymConst{\HOLTokenStrongEquiv} \HOLConst{relab} \HOLFreeVar{E\sp{\prime}} \HOLBoundVar{rf}
\end{alltt}

Noticed that, the strong bisimulation equivalence is co-inductively
defined, and two processes are strong equivalent if there's a
bisimulation containing them. Thus, to prove two processes are
strong equivalent, it's enough to find a bisimulation containing
them. To prove the they're not strong equivalent, it's enough to try
to construct a bisimulation starting from them and the proof is
finished whenever the
attempt fails. In any case, there's no need to do induction on the
data type of involved CCS processes.

Here are the strong laws proved for the sum operator: (noticed that,
the lack of some parentheses is because we have defined the sum and
parallel operators as left-associative)
\begin{alltt}
STRONG_SUM_IDENT_R:        \HOLTokenTurnstile{} \HOLFreeVar{E} \HOLSymConst{+} \HOLConst{nil} \HOLSymConst{\HOLTokenStrongEquiv} \HOLFreeVar{E}
STRONG_SUM_IDEMP:          \HOLTokenTurnstile{} \HOLFreeVar{E} \HOLSymConst{+} \HOLFreeVar{E} \HOLSymConst{\HOLTokenStrongEquiv} \HOLFreeVar{E}
STRONG_SUM_COMM:           \HOLTokenTurnstile{} \HOLFreeVar{E} \HOLSymConst{+} \HOLFreeVar{E\sp{\prime}} \HOLSymConst{\HOLTokenStrongEquiv} \HOLFreeVar{E\sp{\prime}} \HOLSymConst{+} \HOLFreeVar{E}
STRONG_SUM_IDENT_L:        \HOLTokenTurnstile{} \HOLConst{nil} \HOLSymConst{+} \HOLFreeVar{E} \HOLSymConst{\HOLTokenStrongEquiv} \HOLFreeVar{E}
STRONG_SUM_ASSOC_R:        \HOLTokenTurnstile{} \HOLFreeVar{E} \HOLSymConst{+} \HOLFreeVar{E\sp{\prime}} \HOLSymConst{+} \HOLFreeVar{E\sp{\prime\prime}} \HOLSymConst{\HOLTokenStrongEquiv} \HOLFreeVar{E} \HOLSymConst{+} (\HOLFreeVar{E\sp{\prime}} \HOLSymConst{+} \HOLFreeVar{E\sp{\prime\prime}})
STRONG_SUM_ASSOC_L:        \HOLTokenTurnstile{} \HOLFreeVar{E} \HOLSymConst{+} (\HOLFreeVar{E\sp{\prime}} \HOLSymConst{+} \HOLFreeVar{E\sp{\prime\prime}}) \HOLSymConst{\HOLTokenStrongEquiv} \HOLFreeVar{E} \HOLSymConst{+} \HOLFreeVar{E\sp{\prime}} \HOLSymConst{+} \HOLFreeVar{E\sp{\prime\prime}}
STRONG_SUM_MID_IDEMP:      \HOLTokenTurnstile{} \HOLFreeVar{E} \HOLSymConst{+} \HOLFreeVar{E\sp{\prime}} \HOLSymConst{+} \HOLFreeVar{E} \HOLSymConst{\HOLTokenStrongEquiv} \HOLFreeVar{E\sp{\prime}} \HOLSymConst{+} \HOLFreeVar{E}
STRONG_LEFT_SUM_MID_IDEMP: \HOLTokenTurnstile{} \HOLFreeVar{E} \HOLSymConst{+} \HOLFreeVar{E\sp{\prime}} \HOLSymConst{+} \HOLFreeVar{E\sp{\prime\prime}} \HOLSymConst{+} \HOLFreeVar{E\sp{\prime}} \HOLSymConst{\HOLTokenStrongEquiv} \HOLFreeVar{E} \HOLSymConst{+} \HOLFreeVar{E\sp{\prime\prime}} \HOLSymConst{+} \HOLFreeVar{E\sp{\prime}}
\end{alltt}
Not all proven theorems are fundamental (in the sense of providing a
minimal axiomatization set for proving all other algebraic laws). The
first several theorems must be proved by constructing bisimulation
relations and then verifying the definitions of strong bisimulation
and strong equivalence, and their formal proofs were written in
goal-directed ways. Instead, the
last three ones were all constructed in forward way by applications of
previous proven algebraic laws, without directly using any SOS
inference rules and the definition of strong equivalence. Such
constructions were based on two useful ML functions \texttt{S_SYM} and
\texttt{S_TRANS} which builds new strong laws from the symmetry and
transitivity of strong equivalence:
\begin{lstlisting}
(* Define S_SYM such that, when given a theorem A |- STRONG_EQUIV t1 t2,
   returns the theorem A |- STRONG_EQUIV t2 t1. *)
fun S_SYM thm = MATCH_MP STRONG_EQUIV_SYM thm;

(* Define S_TRANS such that, when given the theorems thm1 and thm2, applies
   STRONG_EQUIV_TRANS on them, if possible. *)
fun S_TRANS thm1 thm2 =
    if rhs_tm thm1 = lhs_tm thm2 then
       MATCH_MP STRONG_EQUIV_TRANS (CONJ thm1 thm2)
    else
       failwith "transitivity of strong equivalence not applicable";
\end{lstlisting}
For instance, to construct the proof of \texttt{STRONG_SUM_MID_IDEMP},
the following code was written:
\begin{lstlisting}
(* STRONG_SUM_MID_IDEMP:
   |- !E E'. STRONG_EQUIV (sum (sum E E') E) (sum E' E)
 *)
val STRONG_SUM_MID_IDEMP = save_thm (
   "STRONG_SUM_MID_IDEMP",
    GEN ``E: CCS``
     (GEN ``E': CCS``
       (S_TRANS
        (SPEC ``E: CCS``
         (MATCH_MP STRONG_EQUIV_SUBST_SUM_R
          (SPECL [``E: CCS``, ``E': CCS``] STRONG_SUM_COMM)))
        (S_TRANS
         (SPECL [``E': CCS``, ``E: CCS``, ``E: CCS``] STRONG_SUM_ASSOC_R)
         (SPEC ``E': CCS``
          (MATCH_MP STRONG_EQUIV_SUBST_SUM_L
           (SPEC ``E: CCS`` STRONG_SUM_IDEMP)))))));
\end{lstlisting}

Here are the strong laws we have proved for the par operator:
\begin{alltt}
STRONG_PAR_IDENT_R:        \HOLTokenTurnstile{} \HOLFreeVar{E} \HOLSymConst{||} \HOLConst{nil} \HOLSymConst{\HOLTokenStrongEquiv} \HOLFreeVar{E}
STRONG_PAR_COMM:           \HOLTokenTurnstile{} \HOLFreeVar{E} \HOLSymConst{||} \HOLFreeVar{E\sp{\prime}} \HOLSymConst{\HOLTokenStrongEquiv} \HOLFreeVar{E\sp{\prime}} \HOLSymConst{||} \HOLFreeVar{E}
STRONG_PAR_IDENT_L:        \HOLTokenTurnstile{} \HOLConst{nil} \HOLSymConst{||} \HOLFreeVar{E} \HOLSymConst{\HOLTokenStrongEquiv} \HOLFreeVar{E}
STRONG_PAR_ASSOC:          \HOLTokenTurnstile{} \HOLFreeVar{E} \HOLSymConst{||} \HOLFreeVar{E\sp{\prime}} \HOLSymConst{||} \HOLFreeVar{E\sp{\prime\prime}} \HOLSymConst{\HOLTokenStrongEquiv} \HOLFreeVar{E} \HOLSymConst{||} (\HOLFreeVar{E\sp{\prime}} \HOLSymConst{||} \HOLFreeVar{E\sp{\prime\prime}})
STRONG_PAR_PREF_TAU:       \HOLTokenTurnstile{} \HOLFreeVar{u}\HOLSymConst{..}\HOLFreeVar{E} \HOLSymConst{||} \HOLSymConst{\ensuremath{\tau}}\HOLSymConst{..}\HOLFreeVar{E\sp{\prime}} \HOLSymConst{\HOLTokenStrongEquiv} \HOLFreeVar{u}\HOLSymConst{..}(\HOLFreeVar{E} \HOLSymConst{||} \HOLSymConst{\ensuremath{\tau}}\HOLSymConst{..}\HOLFreeVar{E\sp{\prime}}) \HOLSymConst{+} \HOLSymConst{\ensuremath{\tau}}\HOLSymConst{..}(\HOLFreeVar{u}\HOLSymConst{..}\HOLFreeVar{E} \HOLSymConst{||} \HOLFreeVar{E\sp{\prime}})
STRONG_PAR_TAU_PREF:       \HOLTokenTurnstile{} \HOLSymConst{\ensuremath{\tau}}\HOLSymConst{..}\HOLFreeVar{E} \HOLSymConst{||} \HOLFreeVar{u}\HOLSymConst{..}\HOLFreeVar{E\sp{\prime}} \HOLSymConst{\HOLTokenStrongEquiv} \HOLSymConst{\ensuremath{\tau}}\HOLSymConst{..}(\HOLFreeVar{E} \HOLSymConst{||} \HOLFreeVar{u}\HOLSymConst{..}\HOLFreeVar{E\sp{\prime}}) \HOLSymConst{+} \HOLFreeVar{u}\HOLSymConst{..}(\HOLSymConst{\ensuremath{\tau}}\HOLSymConst{..}\HOLFreeVar{E} \HOLSymConst{||} \HOLFreeVar{E\sp{\prime}})
STRONG_PAR_TAU_TAU:        \HOLTokenTurnstile{} \HOLSymConst{\ensuremath{\tau}}\HOLSymConst{..}\HOLFreeVar{E} \HOLSymConst{||} \HOLSymConst{\ensuremath{\tau}}\HOLSymConst{..}\HOLFreeVar{E\sp{\prime}} \HOLSymConst{\HOLTokenStrongEquiv} \HOLSymConst{\ensuremath{\tau}}\HOLSymConst{..}(\HOLFreeVar{E} \HOLSymConst{||} \HOLSymConst{\ensuremath{\tau}}\HOLSymConst{..}\HOLFreeVar{E\sp{\prime}}) \HOLSymConst{+} \HOLSymConst{\ensuremath{\tau}}\HOLSymConst{..}(\HOLSymConst{\ensuremath{\tau}}\HOLSymConst{..}\HOLFreeVar{E} \HOLSymConst{||} \HOLFreeVar{E\sp{\prime}})
STRONG_PAR_PREF_NO_SYNCR:
\HOLTokenTurnstile{} \HOLFreeVar{l} \HOLSymConst{\HOLTokenNotEqual{}} \HOLConst{COMPL} \HOLFreeVar{l\sp{\prime}} \HOLSymConst{\HOLTokenImp{}}
   \HOLSymConst{\HOLTokenForall{}}\HOLBoundVar{E} \HOLBoundVar{E\sp{\prime}}.
     \HOLConst{label} \HOLFreeVar{l}\HOLSymConst{..}\HOLBoundVar{E} \HOLSymConst{||} \HOLConst{label} \HOLFreeVar{l\sp{\prime}}\HOLSymConst{..}\HOLBoundVar{E\sp{\prime}} \HOLSymConst{\HOLTokenStrongEquiv}
     \HOLConst{label} \HOLFreeVar{l}\HOLSymConst{..}(\HOLBoundVar{E} \HOLSymConst{||} \HOLConst{label} \HOLFreeVar{l\sp{\prime}}\HOLSymConst{..}\HOLBoundVar{E\sp{\prime}}) \HOLSymConst{+}
     \HOLConst{label} \HOLFreeVar{l\sp{\prime}}\HOLSymConst{..}(\HOLConst{label} \HOLFreeVar{l}\HOLSymConst{..}\HOLBoundVar{E} \HOLSymConst{||} \HOLBoundVar{E\sp{\prime}})

STRONG_PAR_PREF_SYNCR:
\HOLTokenTurnstile{} (\HOLFreeVar{l} \HOLSymConst{=} \HOLConst{COMPL} \HOLFreeVar{l\sp{\prime}}) \HOLSymConst{\HOLTokenImp{}}
   \HOLSymConst{\HOLTokenForall{}}\HOLBoundVar{E} \HOLBoundVar{E\sp{\prime}}.
     \HOLConst{label} \HOLFreeVar{l}\HOLSymConst{..}\HOLBoundVar{E} \HOLSymConst{||} \HOLConst{label} \HOLFreeVar{l\sp{\prime}}\HOLSymConst{..}\HOLBoundVar{E\sp{\prime}} \HOLSymConst{\HOLTokenStrongEquiv}
     \HOLConst{label} \HOLFreeVar{l}\HOLSymConst{..}(\HOLBoundVar{E} \HOLSymConst{||} \HOLConst{label} \HOLFreeVar{l\sp{\prime}}\HOLSymConst{..}\HOLBoundVar{E\sp{\prime}}) \HOLSymConst{+}
     \HOLConst{label} \HOLFreeVar{l\sp{\prime}}\HOLSymConst{..}(\HOLConst{label} \HOLFreeVar{l}\HOLSymConst{..}\HOLBoundVar{E} \HOLSymConst{||} \HOLBoundVar{E\sp{\prime}}) \HOLSymConst{+} \HOLSymConst{\ensuremath{\tau}}\HOLSymConst{..}(\HOLBoundVar{E} \HOLSymConst{||} \HOLBoundVar{E\sp{\prime}})
\end{alltt}

And the strong laws for the restriction operator:
\begin{alltt}
STRONG_RESTR_NIL:          \HOLTokenTurnstile{} \HOLSymConst{\ensuremath{\nu}} \HOLFreeVar{L} \HOLConst{nil} \HOLSymConst{\HOLTokenStrongEquiv} \HOLConst{nil}
STRONG_RESTR_SUM:          \HOLTokenTurnstile{} \HOLSymConst{\ensuremath{\nu}} \HOLFreeVar{L} (\HOLFreeVar{E} \HOLSymConst{+} \HOLFreeVar{E\sp{\prime}}) \HOLSymConst{\HOLTokenStrongEquiv} \HOLSymConst{\ensuremath{\nu}} \HOLFreeVar{L} \HOLFreeVar{E} \HOLSymConst{+} \HOLSymConst{\ensuremath{\nu}} \HOLFreeVar{L} \HOLFreeVar{E\sp{\prime}}
STRONG_RESTR_PREFIX_TAU:   \HOLTokenTurnstile{} \HOLSymConst{\ensuremath{\nu}} \HOLFreeVar{L} (\HOLSymConst{\ensuremath{\tau}}\HOLSymConst{..}\HOLFreeVar{E}) \HOLSymConst{\HOLTokenStrongEquiv} \HOLSymConst{\ensuremath{\tau}}\HOLSymConst{..}\HOLSymConst{\ensuremath{\nu}} \HOLFreeVar{L} \HOLFreeVar{E}
STRONG_RESTR_PR_LAB_NIL:   \HOLTokenTurnstile{} \HOLFreeVar{l} \HOLSymConst{\HOLTokenIn{}} \HOLFreeVar{L} \HOLSymConst{\HOLTokenDisj{}} \HOLConst{COMPL} \HOLFreeVar{l} \HOLSymConst{\HOLTokenIn{}} \HOLFreeVar{L} \HOLSymConst{\HOLTokenImp{}} \HOLSymConst{\HOLTokenForall{}}\HOLBoundVar{E}. \HOLSymConst{\ensuremath{\nu}} \HOLFreeVar{L} (\HOLConst{label} \HOLFreeVar{l}\HOLSymConst{..}\HOLBoundVar{E}) \HOLSymConst{\HOLTokenStrongEquiv} \HOLConst{nil}
STRONG_RESTR_PREFIX_LABEL: \HOLTokenTurnstile{} \HOLFreeVar{l} \HOLSymConst{\HOLTokenNotIn{}} \HOLFreeVar{L} \HOLSymConst{\HOLTokenConj{}} \HOLConst{COMPL} \HOLFreeVar{l} \HOLSymConst{\HOLTokenNotIn{}} \HOLFreeVar{L} \HOLSymConst{\HOLTokenImp{}}
   \HOLSymConst{\HOLTokenForall{}}\HOLBoundVar{E}. \HOLSymConst{\ensuremath{\nu}} \HOLFreeVar{L} (\HOLConst{label} \HOLFreeVar{l}\HOLSymConst{..}\HOLBoundVar{E}) \HOLSymConst{\HOLTokenStrongEquiv} \HOLConst{label} \HOLFreeVar{l}\HOLSymConst{..}\HOLSymConst{\ensuremath{\nu}} \HOLFreeVar{L} \HOLBoundVar{E}
\end{alltt}

The strong laws for the relabeling operator:
\begin{alltt}
STRONG_RELAB_NIL:          \HOLTokenTurnstile{} \HOLConst{relab} \HOLConst{nil} \HOLFreeVar{rf} \HOLSymConst{\HOLTokenStrongEquiv} \HOLConst{nil}
STRONG_RELAB_SUM:          \HOLTokenTurnstile{} \HOLConst{relab} (\HOLFreeVar{E} \HOLSymConst{+} \HOLFreeVar{E\sp{\prime}}) \HOLFreeVar{rf} \HOLSymConst{\HOLTokenStrongEquiv} \HOLConst{relab} \HOLFreeVar{E} \HOLFreeVar{rf} \HOLSymConst{+} \HOLConst{relab} \HOLFreeVar{E\sp{\prime}} \HOLFreeVar{rf}
STRONG_RELAB_PREFIX:       \HOLTokenTurnstile{} \HOLConst{relab} (\HOLFreeVar{u}\HOLSymConst{..}\HOLFreeVar{E}) (\HOLConst{RELAB} \HOLFreeVar{labl}) \HOLSymConst{\HOLTokenStrongEquiv}
   \HOLConst{relabel} (\HOLConst{RELAB} \HOLFreeVar{labl}) \HOLFreeVar{u}\HOLSymConst{..}\HOLConst{relab} \HOLFreeVar{E} (\HOLConst{RELAB} \HOLFreeVar{labl})
\end{alltt}

The strong laws for the recursion operator (for constants):
\begin{alltt}
STRONG_UNFOLDING:          \HOLTokenTurnstile{} \HOLConst{rec} \HOLFreeVar{X} \HOLFreeVar{E} \HOLSymConst{\HOLTokenStrongEquiv} \HOLConst{CCS_Subst} \HOLFreeVar{E} (\HOLConst{rec} \HOLFreeVar{X} \HOLFreeVar{E}) \HOLFreeVar{X}
STRONG_PREF_REC_EQUIV:     \HOLTokenTurnstile{} \HOLFreeVar{u}\HOLSymConst{..}\HOLConst{rec} \HOLFreeVar{s} (\HOLFreeVar{v}\HOLSymConst{..}\HOLFreeVar{u}\HOLSymConst{..}\HOLConst{var} \HOLFreeVar{s}) \HOLSymConst{\HOLTokenStrongEquiv} \HOLConst{rec} \HOLFreeVar{s} (\HOLFreeVar{u}\HOLSymConst{..}\HOLFreeVar{v}\HOLSymConst{..}\HOLConst{var} \HOLFreeVar{s})
STRONG_REC_ACT2:           \HOLTokenTurnstile{} \HOLConst{rec} \HOLFreeVar{s} (\HOLFreeVar{u}\HOLSymConst{..}\HOLFreeVar{u}\HOLSymConst{..}\HOLConst{var} \HOLFreeVar{s}) \HOLSymConst{\HOLTokenStrongEquiv} \HOLConst{rec} \HOLFreeVar{s} (\HOLFreeVar{u}\HOLSymConst{..}\HOLConst{var} \HOLFreeVar{s})
\end{alltt}
All above three theorems for recursion operator were fundamental (in the sense that, they
cannot be proved by just using other strong laws).

Finally, all above strong laws could be used either manually or as part of the
decision procedure for automatically deciding strong equivalences
between two CCS process. However such a decision procedure is not done
in the current project.

\subsection{Expansion Law for strong equivalence}

The final big piece of proof work in this project is the
representation and proof of the following \emph{expansion law} (sometimes
also called the \emph{interleaving law}:
\begin{proposition}{(Expansion Law)}
Let $p = \sum_{i=1}^n \mu_i.p_i$ and $q = \sum_{j=1}^m
\mu'_j.q_j$. Then
\begin{equation}
p | q \sim \sum_{i=1}^n \mu_i.(p_i | q) + \sum_{j=1}^m \mu'_j.(p|q_j)
+ \sum_{i,j:\overline{\mu_i}=\mu'_j} \tau.(p_i | q_j)
\end{equation}
\end{proposition}

Some characteristics made the formal proof very special and different from all other theorems that
we have proved so far. First of all, arithmetic numbers (of type \HOLinline{\HOLTyOp{num}}) were involved for the first
  time, and now our CCS theory depends on elementary mathematical theories provided by HOL,
  namely the \texttt{prim_recTheory} and
  \texttt{arithmeticTheory}. Although arithmetic operations like $+,
  -, \cdot, /$ were not involved (yet), but we do need to compare
  number values and use some related theorems.

Also two CCS accessors were defined and used to access the internal
  structure of CCS processes, namely \HOLinline{\HOLConst{PREF_ACT}} for getting the
  initial action and \HOLinline{\HOLConst{PREF_PROC}} for getting the rest of process
  without the first action. Together there's predicate
  \HOLinline{\HOLConst{Is_Prefix}} for testing if a CCS is a prefixed process:
\begin{alltt}
\HOLTokenTurnstile{} \HOLConst{PREF_ACT} (\HOLFreeVar{u}\HOLSymConst{..}\HOLFreeVar{E}) \HOLSymConst{=} \HOLFreeVar{u}
\HOLTokenTurnstile{} \HOLConst{PREF_PROC} (\HOLFreeVar{u}\HOLSymConst{..}\HOLFreeVar{E}) \HOLSymConst{=} \HOLFreeVar{E}
\HOLTokenTurnstile{} \HOLConst{Is_Prefix} \HOLFreeVar{E} \HOLSymConst{\HOLTokenEquiv{}} \HOLSymConst{\HOLTokenExists{}}\HOLBoundVar{u} \HOLBoundVar{E\sp{\prime}}. \HOLFreeVar{E} \HOLSymConst{=} \HOLBoundVar{u}\HOLSymConst{..}\HOLBoundVar{E\sp{\prime}}
\end{alltt}
They are needed because we're going to represent $\mu_i.p_i$ as the
value of a function: $f(i)$
in which $f$ has the type \HOLinline{\HOLTyOp{num} -> \HOLTyOp{CCS}}. And in
this way, to get $\mu_i$ and $p_i$ we have to use accessors:
``\HOLinline{\HOLConst{PREF_ACT} (\HOLFreeVar{f} \HOLFreeVar{i})}'' and ``\HOLinline{\HOLConst{PREF_PROC} (\HOLFreeVar{f} \HOLFreeVar{i})}''.

The next job is to represent a finite sum of CCS processes. This is
done by the following recursive function \HOLinline{\HOLConst{SIGMA}}:
\begin{alltt}
\HOLTokenTurnstile{} (\HOLSymConst{\HOLTokenForall{}}\HOLBoundVar{f}. \HOLConst{SIGMA} \HOLBoundVar{f} \HOLNumLit{0} \HOLSymConst{=} \HOLBoundVar{f} \HOLNumLit{0}) \HOLSymConst{\HOLTokenConj{}}
   \HOLSymConst{\HOLTokenForall{}}\HOLBoundVar{f} \HOLBoundVar{n}. \HOLConst{SIGMA} \HOLBoundVar{f} (\HOLConst{SUC} \HOLBoundVar{n}) \HOLSymConst{=} \HOLConst{SIGMA} \HOLBoundVar{f} \HOLBoundVar{n} \HOLSymConst{+} \HOLBoundVar{f} (\HOLConst{SUC} \HOLBoundVar{n})
\end{alltt}
Thus if there's a function $f$ of type \HOLinline{\HOLTyOp{num} -> \HOLTyOp{CCS}}, we should
be able to represent $\sum_{i=1}^n f(i)$ by HOL term
``\HOLinline{\HOLConst{SIGMA} \HOLFreeVar{f} \HOLFreeVar{n}}''.

Now if we took a deeper look at the last summation of the right side
of the expansion law, i.e. $\sum_{i,j:\overline{\mu_i}=\mu'_j}
\tau.(p_i | q_j)$, we found that such a ``sum'' cannot be represented
directly, because there're two index $i,j$ and their possible value
pairs used in the sum depends on the synchronization of corresponding
actions from each $p_i$ and $q_j$.  What we actually need is a
recursively defined function taking all the $p_i$ and $q_j$ and return
the synchronized process in forms like $\sum \tau.(p_i | q_j)$.

But this is still too complicated, instead we first define functions to synchronize
just one process with another group of processes. This work is
achieved by the function \HOLinline{\HOLConst{SYNC}} of type \HOLinline{\HOLTyOp{Action} -> \HOLTyOp{CCS} -> (\HOLTyOp{num} -> \HOLTyOp{CCS}) -> \HOLTyOp{num} -> \HOLTyOp{CCS}}:
\begin{alltt}
\HOLTokenTurnstile{} (\HOLSymConst{\HOLTokenForall{}}\HOLBoundVar{u} \HOLBoundVar{P} \HOLBoundVar{f}.
      \HOLConst{SYNC} \HOLBoundVar{u} \HOLBoundVar{P} \HOLBoundVar{f} \HOLNumLit{0} \HOLSymConst{=}
      \HOLKeyword{if} (\HOLBoundVar{u} \HOLSymConst{=} \HOLSymConst{\ensuremath{\tau}}) \HOLSymConst{\HOLTokenDisj{}} (\HOLConst{PREF_ACT} (\HOLBoundVar{f} \HOLNumLit{0}) \HOLSymConst{=} \HOLSymConst{\ensuremath{\tau}}) \HOLKeyword{then} \HOLConst{nil}
      \HOLKeyword{else} \HOLKeyword{if} \HOLConst{LABEL} \HOLBoundVar{u} \HOLSymConst{=} \HOLConst{COMPL} (\HOLConst{LABEL} (\HOLConst{PREF_ACT} (\HOLBoundVar{f} \HOLNumLit{0}))) \HOLKeyword{then}
        \HOLSymConst{\ensuremath{\tau}}\HOLSymConst{..}(\HOLBoundVar{P} \HOLSymConst{||} \HOLConst{PREF_PROC} (\HOLBoundVar{f} \HOLNumLit{0}))
      \HOLKeyword{else} \HOLConst{nil}) \HOLSymConst{\HOLTokenConj{}}
   \HOLSymConst{\HOLTokenForall{}}\HOLBoundVar{u} \HOLBoundVar{P} \HOLBoundVar{f} \HOLBoundVar{n}.
     \HOLConst{SYNC} \HOLBoundVar{u} \HOLBoundVar{P} \HOLBoundVar{f} (\HOLConst{SUC} \HOLBoundVar{n}) \HOLSymConst{=}
     \HOLKeyword{if} (\HOLBoundVar{u} \HOLSymConst{=} \HOLSymConst{\ensuremath{\tau}}) \HOLSymConst{\HOLTokenDisj{}} (\HOLConst{PREF_ACT} (\HOLBoundVar{f} (\HOLConst{SUC} \HOLBoundVar{n})) \HOLSymConst{=} \HOLSymConst{\ensuremath{\tau}}) \HOLKeyword{then}
       \HOLConst{SYNC} \HOLBoundVar{u} \HOLBoundVar{P} \HOLBoundVar{f} \HOLBoundVar{n}
     \HOLKeyword{else} \HOLKeyword{if}
       \HOLConst{LABEL} \HOLBoundVar{u} \HOLSymConst{=} \HOLConst{COMPL} (\HOLConst{LABEL} (\HOLConst{PREF_ACT} (\HOLBoundVar{f} (\HOLConst{SUC} \HOLBoundVar{n}))))
     \HOLKeyword{then}
       \HOLSymConst{\ensuremath{\tau}}\HOLSymConst{..}(\HOLBoundVar{P} \HOLSymConst{||} \HOLConst{PREF_PROC} (\HOLBoundVar{f} (\HOLConst{SUC} \HOLBoundVar{n}))) \HOLSymConst{+} \HOLConst{SYNC} \HOLBoundVar{u} \HOLBoundVar{P} \HOLBoundVar{f} \HOLBoundVar{n}
     \HOLKeyword{else} \HOLConst{SYNC} \HOLBoundVar{u} \HOLBoundVar{P} \HOLBoundVar{f} \HOLBoundVar{n}
\end{alltt}

Then the synchronization of two group of processes can be further
defined by another recursive function \HOLinline{\HOLConst{ALL_SYNC}} of type
\HOLinline{(\HOLTyOp{num} -> \HOLTyOp{CCS}) -> \HOLTyOp{num} -> (\HOLTyOp{num} -> \HOLTyOp{CCS}) -> \HOLTyOp{num} -> \HOLTyOp{CCS}}:
\begin{alltt}
\HOLTokenTurnstile{} (\HOLSymConst{\HOLTokenForall{}}\HOLBoundVar{f} \HOLBoundVar{f\sp{\prime}} \HOLBoundVar{m}.
      \HOLConst{ALL_SYNC} \HOLBoundVar{f} \HOLNumLit{0} \HOLBoundVar{f\sp{\prime}} \HOLBoundVar{m} \HOLSymConst{=}
      \HOLConst{SYNC} (\HOLConst{PREF_ACT} (\HOLBoundVar{f} \HOLNumLit{0})) (\HOLConst{PREF_PROC} (\HOLBoundVar{f} \HOLNumLit{0})) \HOLBoundVar{f\sp{\prime}} \HOLBoundVar{m}) \HOLSymConst{\HOLTokenConj{}}
   \HOLSymConst{\HOLTokenForall{}}\HOLBoundVar{f} \HOLBoundVar{n} \HOLBoundVar{f\sp{\prime}} \HOLBoundVar{m}.
     \HOLConst{ALL_SYNC} \HOLBoundVar{f} (\HOLConst{SUC} \HOLBoundVar{n}) \HOLBoundVar{f\sp{\prime}} \HOLBoundVar{m} \HOLSymConst{=}
     \HOLConst{ALL_SYNC} \HOLBoundVar{f} \HOLBoundVar{n} \HOLBoundVar{f\sp{\prime}} \HOLBoundVar{m} \HOLSymConst{+}
     \HOLConst{SYNC} (\HOLConst{PREF_ACT} (\HOLBoundVar{f} (\HOLConst{SUC} \HOLBoundVar{n}))) (\HOLConst{PREF_PROC} (\HOLBoundVar{f} (\HOLConst{SUC} \HOLBoundVar{n}))) \HOLBoundVar{f\sp{\prime}} \HOLBoundVar{m}
\end{alltt}

Some lemmas about \HOLinline{\HOLConst{SIGMA}} and the two synchronization
functions were proved first:
\begin{alltt}
SIGMA_TRANS_THM_EQ:
\HOLTokenTurnstile{} \HOLConst{SIGMA} \HOLFreeVar{f} \HOLFreeVar{n} --\HOLFreeVar{u}-> \HOLFreeVar{E} \HOLSymConst{\HOLTokenEquiv{}} \HOLSymConst{\HOLTokenExists{}}\HOLBoundVar{k}. \HOLBoundVar{k} \HOLSymConst{\HOLTokenLeq{}} \HOLFreeVar{n} \HOLSymConst{\HOLTokenConj{}} \HOLFreeVar{f} \HOLBoundVar{k} --\HOLFreeVar{u}-> \HOLFreeVar{E}
SYNC_TRANS_THM_EQ:
\HOLTokenTurnstile{} \HOLConst{SYNC} \HOLFreeVar{u} \HOLFreeVar{P} \HOLFreeVar{f} \HOLFreeVar{m} --\HOLFreeVar{v}-> \HOLFreeVar{Q} \HOLSymConst{\HOLTokenEquiv{}}
   \HOLSymConst{\HOLTokenExists{}}\HOLBoundVar{j} \HOLBoundVar{l}.
     \HOLBoundVar{j} \HOLSymConst{\HOLTokenLeq{}} \HOLFreeVar{m} \HOLSymConst{\HOLTokenConj{}} (\HOLFreeVar{u} \HOLSymConst{=} \HOLConst{label} \HOLBoundVar{l}) \HOLSymConst{\HOLTokenConj{}}
     (\HOLConst{PREF_ACT} (\HOLFreeVar{f} \HOLBoundVar{j}) \HOLSymConst{=} \HOLConst{label} (\HOLConst{COMPL} \HOLBoundVar{l})) \HOLSymConst{\HOLTokenConj{}} (\HOLFreeVar{v} \HOLSymConst{=} \HOLSymConst{\ensuremath{\tau}}) \HOLSymConst{\HOLTokenConj{}}
     (\HOLFreeVar{Q} \HOLSymConst{=} \HOLFreeVar{P} \HOLSymConst{||} \HOLConst{PREF_PROC} (\HOLFreeVar{f} \HOLBoundVar{j}))
ALL_SYNC_TRANS_THM_EQ:
\HOLTokenTurnstile{} \HOLConst{ALL_SYNC} \HOLFreeVar{f} \HOLFreeVar{n} \HOLFreeVar{f\sp{\prime}} \HOLFreeVar{m} --\HOLFreeVar{u}-> \HOLFreeVar{E} \HOLSymConst{\HOLTokenEquiv{}}
   \HOLSymConst{\HOLTokenExists{}}\HOLBoundVar{k} \HOLBoundVar{k\sp{\prime}} \HOLBoundVar{l}.
     \HOLBoundVar{k} \HOLSymConst{\HOLTokenLeq{}} \HOLFreeVar{n} \HOLSymConst{\HOLTokenConj{}} \HOLBoundVar{k\sp{\prime}} \HOLSymConst{\HOLTokenLeq{}} \HOLFreeVar{m} \HOLSymConst{\HOLTokenConj{}} (\HOLConst{PREF_ACT} (\HOLFreeVar{f} \HOLBoundVar{k}) \HOLSymConst{=} \HOLConst{label} \HOLBoundVar{l}) \HOLSymConst{\HOLTokenConj{}}
     (\HOLConst{PREF_ACT} (\HOLFreeVar{f\sp{\prime}} \HOLBoundVar{k\sp{\prime}}) \HOLSymConst{=} \HOLConst{label} (\HOLConst{COMPL} \HOLBoundVar{l})) \HOLSymConst{\HOLTokenConj{}} (\HOLFreeVar{u} \HOLSymConst{=} \HOLSymConst{\ensuremath{\tau}}) \HOLSymConst{\HOLTokenConj{}}
     (\HOLFreeVar{E} \HOLSymConst{=} \HOLConst{PREF_PROC} (\HOLFreeVar{f} \HOLBoundVar{k}) \HOLSymConst{||} \HOLConst{PREF_PROC} (\HOLFreeVar{f\sp{\prime}} \HOLBoundVar{k\sp{\prime}}))
\end{alltt}

Finally, we have proved the Expansion Law in the following form:
\begin{alltt}
STRONG_PAR_LAW:
\HOLTokenTurnstile{} (\HOLSymConst{\HOLTokenForall{}}\HOLBoundVar{i}. \HOLBoundVar{i} \HOLSymConst{\HOLTokenLeq{}} \HOLFreeVar{n} \HOLSymConst{\HOLTokenImp{}} \HOLConst{Is_Prefix} (\HOLFreeVar{f} \HOLBoundVar{i})) \HOLSymConst{\HOLTokenConj{}}
   (\HOLSymConst{\HOLTokenForall{}}\HOLBoundVar{j}. \HOLBoundVar{j} \HOLSymConst{\HOLTokenLeq{}} \HOLFreeVar{m} \HOLSymConst{\HOLTokenImp{}} \HOLConst{Is_Prefix} (\HOLFreeVar{f\sp{\prime}} \HOLBoundVar{j})) \HOLSymConst{\HOLTokenImp{}}
   \HOLConst{SIGMA} \HOLFreeVar{f} \HOLFreeVar{n} \HOLSymConst{||} \HOLConst{SIGMA} \HOLFreeVar{f\sp{\prime}} \HOLFreeVar{m} \HOLSymConst{\HOLTokenStrongEquiv}
   \HOLConst{SIGMA} (\HOLTokenLambda{}\HOLBoundVar{i}. \HOLConst{PREF_ACT} (\HOLFreeVar{f} \HOLBoundVar{i})\HOLSymConst{..}(\HOLConst{PREF_PROC} (\HOLFreeVar{f} \HOLBoundVar{i}) \HOLSymConst{||} \HOLConst{SIGMA} \HOLFreeVar{f\sp{\prime}} \HOLFreeVar{m}))
     \HOLFreeVar{n} \HOLSymConst{+}
   \HOLConst{SIGMA} (\HOLTokenLambda{}\HOLBoundVar{j}. \HOLConst{PREF_ACT} (\HOLFreeVar{f\sp{\prime}} \HOLBoundVar{j})\HOLSymConst{..}(\HOLConst{SIGMA} \HOLFreeVar{f} \HOLFreeVar{n} \HOLSymConst{||} \HOLConst{PREF_PROC} (\HOLFreeVar{f\sp{\prime}} \HOLBoundVar{j})))
     \HOLFreeVar{m} \HOLSymConst{+} \HOLConst{ALL_SYNC} \HOLFreeVar{f} \HOLFreeVar{n} \HOLFreeVar{f\sp{\prime}} \HOLFreeVar{m}
\end{alltt}

\section{Missing pieces and Future directions}

The old proof scripts provided by Prof. Nesi do not contain anything
related to weak bisimulation, while these things were talked as major
work in the original paper. We think those proof scripts must have
been unfortunately lost.\footnote{This is not true any
  more. Prof. Monica Nesi still have all these proof scripts, and the
  author is now waiting for these code to continue the rests of
  the porting work.} In our project, due to time limits we only
re-defined the concepts of weak transitions, weak bisimulation, weak
bisimultion equivalence (observation equivalence) and rooted weak
equivalence (observation congruence), but almost didn't prove any
useful results, except for the experiments to show the correctness of
HOL's co-inductive relation definining facility (\texttt{Hol_coreln})
for weak bisimulation equivalences.  Given the fact that, more
practical model checking were done by comparing (rooted) weak
bisimulation equivalences between two CCS processes, our current work
is far from complete.

The other big missing piece is the decision procedure for
automatic checking of strong (and weak) bisimulation
equivalence. There exists some fast algorithms for bisimulation
equivalence checking, they were mostly based on reductions of the
equivalence checking to the so-called ``coarsest relational
partitioning'' problem. (c.f. \cite{Kanellakis:1990ja} and
\cite{Paige:1987du}), some variants (e.g. \cite{Yannakakis:1992fh}) are
suitable for compact representations like CCS, in which the whole
graph is not visible. On the other side, we know Concurrency Workbench
didn't use the most efficient algorithm (c.f. p.13 of
\cite{CLEAVELAND:1993dr}), and this leaves us a room to create a
faster equivalence checking tool, and it runs even inside a theorem prover!

Thus, the author hopes to continue this project with the following possible
direction:
\begin{enumerate}
\item Complete the theory for (rooted) weak equivalence and prove the
  related weak laws.
\item Create decision procedures for bisimulation equivalence checking
  which take two CCS processes and give a theorem about their
  equivalence.
\item Formally prove some deep theorems for bisimulation equivalences,
  e.g. the Hennessy Lemma (c.f. p.176 of \cite{Gorrieri:2015jt})
\end{enumerate}

Finally, any tool is only useful when it's proven to be useful for
resolving practical problems. But so far we haven't shown anything for
its applications. Indeed, maybe we can never show more useful results
than those already have with software like Concurrency Workbench. So
our main hope here, is to provide experiences and good basis for building more
complicated process algebras (e.g. CCS variants with more opeartors). And when future researchers published
new theorems in this area, maybe they could provide also formal proofs
using the framework provided in this project.

\section{On the choice of HOL}

On the initial choice of using HOL for the CCS formalization, we
believe this was partially influenced by the theorem proving
environment in Britain in 1990s, and the fact that, the creator of HOL
(Mike J. C. Gordon) was also working in Cambridge University since
1981.  Students and researchers usually choose to use software developed and
taught by scholars in their own university, or country. The latter
case is particularly true for  researchers of formal methods in France:
they almost always use Coq and OCaml to build everything.

Beside environment reasons, and the fact that the work in this project
is not original (porting old code), the author still thinks that HOL4
is a better choice than other popular theorem provers like Coq and
Isabelle. This is becase, only in HOL4, it's possible to write proof
scripts, new tacticals and ordinary functions (which generates new
theorems) in the same underlying programming language (Standard
ML). And the powerful source-level debugging support in Poly/ML was
essential to us for fixing the bugs found in the big function
\texttt{CCS_TRANS_CONV}.

There's one extra benefit to use theorem prover built on top of
Standard ML: it's a smaller language than OCaml \footnote{Both Matita
  and Coq were written in OCaml, however Coq (and maybe also Matita) didn't use any OO feature
  provided by OCaml.},
 and currently there's
an ongoing project called CakeML\footnote{\url{https://cakeml.org}}
with the aim to formally verify the compiler for the substantial
subset of Standard ML. When this project finally succeeds to build HOL
on top of it, we'll have a fully trusted computing environment
including the underlying programming language, the theorem prover and
the formal theories. With other programming languages and theorem
provers written in them
, there's little hope in short future to achieve the same level of confidence.

On the other side, Matita\footnote{\url{http://matita.cs.unibo.it/}} was written in OCaml,
a programming language more complex than Standard ML.
Matita is a pure graphics program, and there's no interactive interface.
It lacks of a rich theorem library (there's even no RTC support in its relation theory) that users could benefit.
And there's no interface to develop new tacticals and other proof tools directly at OCaml level.
Due to these limitations, Matita can't be used in projects like the current one. Not to mention that,
Matita is not actively developed any more, and its user group can be almost ignored.

From the view of the author, HOL4 has been designed in an unique way that no other theorem prover
can provide the same feature sets.

\section{Conclusions}

In this exam paper (and project report), we have successfully ported
the old formalization of process algebra CCS (no value passing,
with explicit relabeling operator) from HOL88 to latest HOL4
(Kananaskis-11 and later). We started from the definition and syntax
of CCS processes defined as inductive datatypes in Higher Order Logic,
and then defined all the SOS (Structured Operational Semantics)
inference rules as an relation \texttt{TRANS}. Then all the algebraic
laws including the Expansion Theorem were proved on top of CCS
datatype and SOS rules.

The other big work in this project is a single  ML function (part of
the old work, but we have fixed and enhanced the code) which could automatically compute
the possible transitions for a given CCS process. Different from the
similar facility in softwares like Concurrency Workbench, the output of
our function is a theorem. This is kind of trusting computing, as the
only way to build theorems is to construct it from other
theorems. Although the correctness of such a program is not formally
verified, but as long as it terminates with a theorem as output, the output MUST
be correct.  We have future plans to create a similar tool for
equivalence checking, in which equivalence results are theorems
construced from existing manually proved theorems.

This work is based on old CCS formalization in HOL88, done by
Prof. Monica Nesi (of University of L'Aquila, Italy) in 1992-1995 when he was studying at University of Cambridge. Thanks to Prof. Nesi for finding and
sending the old HOL88 proof scripts to the author.

Thanks to Prof. Roberto Gorrieri, who taught CCS and LTS theory to the
author, and his supports on continuing this HOL-CCS project as exam
project of his course.

Thanks to Prof. Andrea Asperti, who taught the interactive theorem proving
techniques to the author, although it's in another different theorem
prover (Matita).

Thanks to people from HOL community (Thomas Tuerk, Michael Norrish,
Ramana Kumar and many others) for resolving issues and doubts the
author met when using HOL theorem prover.

The paper is written in \LaTeX and LNCS template, with theorems
generated automatically by HOL's \TeX
exporting module (\texttt{EmitTex}) from the proof scripts.

\bibliography{hol-ccs}{}
\bibliographystyle{splncs}

\section*{Appendix A: running the proof scripts in HOL4}

Suppose HOL (Kananaskis-11) has been installed%
\footnote{For installation instructions of HOL4, see
  \texttt{https://hol-theorem-prover.org/\#get}. To run the scripts
  mentioned in this paper correctly, please use kananaskis-11 (latest
  released version).} and the two entry
commands \texttt{hol} and \texttt{Holmake} have been made available in
current Shell environment (e.g. their containing directory is in
\texttt{PATH} environment variable).  Copying all above 4 files into a
empty directory and execute \texttt{Holmake}, they should be compiled
correctly with some extra files generated.

Then after executing \texttt{hol}, enter the following commands to
load all CCS related scripts:
\begin{lstlisting}
---------------------------------------------------------------------
       HOL-4 [Kananaskis 11 (logknl, built Sat Apr 29 12:55:33 2017)]

       For introductory HOL help, type: help "hol";
       To exit type <Control>-D
---------------------------------------------------------------------
> load "ExampleTheory";
val it = (): unit
> open ExampleTheory;
...
> 
\end{lstlisting}
Then we can either access already proved theorems storing into ML variables:
\begin{lstlisting}
> STRONG_PAR_LAW;
val it =
   |- !f n f' m.
     (!i. i <= n ==> Is_Prefix (f i)) /\ (!j. j <= m ==> Is_Prefix (f' j)) ==>
     SIGMA f n || SIGMA f' m
 ~
     SIGMA (\i. PREF_ACT (f i)..(PREF_PROC (f i) || SIGMA f' m)) n +
     SIGMA (\j. PREF_ACT (f' j)..(SIGMA f n || PREF_PROC (f' j))) m +
     ALL_SYNC f n f' m:
   thm
\end{lstlisting}

or compute CCS transitions from any given process:
\begin{lstlisting}
> CCS_TRANS ``(nu "a") (In "a"..nil || Out "a"..nil)``;
val it =
   (|- !u E.
     'm "a" (In "a"..nil || Out "a"..nil) --u-> E <=>
     (u = tau) /\ (E = 'm "a" (nil || nil)),
    [(``tau``,
      ``'m "a" (nil || nil)``)]):
   thm * (term * term) list

> CCS_TRANS ``(In "a"..nil || Out "a"..nil)``;
val it =
   (|- !u E.
     In "a"..nil || Out "a"..nil --u-> E <=>
     ((u = In "a") /\ (E = nil || Out "a"..nil) \/
      (u = Out "a") /\ (E = In "a"..nil || nil)) \/
     (u = tau) /\ (E = nil || nil),
    [(``In "a"``,
      ``nil || Out "a"..nil``),
     (``Out "a"``,
      ``In "a"..nil || nil``),
     (``tau``,
      ``nil || nil``)]):
   thm * (term * term) list
\end{lstlisting}

The generated files \texttt{*.sig} contain lists of all proved theorems, the HTML versions were also
generated as \texttt{*.html} files.

\end{document}